\documentclass[10pt, a4paper]{article}

\usepackage{bm} 

\usepackage{enumerate} 

\usepackage[utf8]{inputenc}
\usepackage[T1]{fontenc}

\usepackage{amsmath,amssymb}
\usepackage{mathtools}  
\usepackage{amsthm} 
\usepackage{mathrsfs} 
\usepackage{bbm} 
\usepackage{hyperref}

\usepackage{soul} 

\usepackage{tikz}
\usepackage{tikz-cd} 
\usetikzlibrary{arrows}
\tikzcdset{arrow style=tikz, diagrams={>=stealth}} 
\usepackage{graphicx}  

\usepackage{simplewick}

\usepackage{etoolbox}

\usepackage{amssymb}  
\usepackage{bbold} 

\theoremstyle{plain}

\newtheorem*{theorem*}{Theorem}

\theoremstyle{definition}

\theoremstyle{definition}

\theoremstyle{definition}
\newtheorem*{definition*}{Definition}

\theoremstyle{definition}

\theoremstyle{definition}
\newtheorem*{remark*}{Remark}
\theoremstyle{definition}

\theoremstyle{definition}
\newtheorem*{remarks*}{Remarks}

\theoremstyle{remark}

\theoremstyle{remark}

\newtheoremstyle{rudin-style-generic}
{}
{}
{}
{}
{\bfseries}
{}
{ }
{\thmnumber{#2} \thmname{#1.}\thmnote{\phantom{,}#3 }}
\newtheoremstyle{rudin-style-generic*}{}{}{}{}{\bfseries}{}{ }{\thmname{#1}\thmnote{#3}}
\theoremstyle{rudin-style-generic}
\newtheorem{parpar}{}[section]
\theoremstyle{rudin-style-generic*}
\newtheorem{parpar*}{}[]{}
\newenvironment{parpar-noproof}
{
  \pushQED{\qed}\begin{parpar}}
  {\popQED\end{parpar}}

\newtheoremstyle{rudin-style-theorem}{}{}{\itshape}{}{\bfseries}{}{ }{\thmnumber{\S\hspace{2pt}#2} \thmname{#1.}\thmnote{\phantom{,}#3}}
\theoremstyle{rudin-style-theorem}
\newtheorem{parpar-theorem}[parpar]{Theorem}
\newtheorem{parpar-proposition}[parpar]{Proposition}
\newtheorem{parpar-lemma}[parpar]{Lemma}
\newtheorem{parpar-corollary}[parpar]{Corollary}

\newenvironment{parpar-proof}[1][\scshape    
\proofname]
{%
  \proof[#1]%
  \setlength{\leftskip}{3em}%
}
{%
  \endproof%
}

\usepackage{xparse}
\NewDocumentCommand{\movedownsub}{e{^_}}{%
  \IfNoValueTF{#1}{%
    \IfNoValueF{#2}{^{}}
  }{%
    ^{#1}
  }%
  \IfNoValueF{#2}{_{#2}}
}
\NewDocumentCommand{\latexchi}{}{\chi\movedownsub}

\AtBeginEnvironment{smallproof}{\footnotesize}   

\AtBeginEnvironment{smallerparagraph}{\small}    

\usepackage{mathtools}

\newcommand{\deq}{\stackrel{\mathrm{def}}{=}}
\newcommand{\dd}{\,\mathrm{d}}
\newcommand{\ii}{\mathrm{i}}
\newcommand{\id}{\,\mathbb{I}}

\newcommand{\PP}{\mathbb{P}}
\newcommand{\NN}{\mathbb{N}}

\newcommand{\RR}{\mathbb{R}}
\newcommand{\CC}{\mathbb{C}}
\newcommand{\KK}{\mathbb{K}}

\newcommand{\Gg}{{\mathbf{G}}}
\newcommand{\Hg}{{\mathbf{H}}}
\newcommand{\Ng}{{\mathbf{N}}}
\newcommand{\Sg}{{\mathbf{S}}}

\DeclareMathOperator{\tr}{Tr}

\renewcommand{\st}{\,\mathbf:\,}

\newcommand*{\Cliff}{\mathcal C\ell}
\newcommand*{\CCliff}{\CC\ell}
\newcommand*{\SO}{\mathbf{SO}}
\newcommand*{\SU}{\mathbf{SU}}

\newcommand*{\Spin}{\mathbf{Spin}}
\newcommand*{\SL}{\mathbf{SL}}

\newcommand*{\ISpin}{\mathbf{ISpin}}

\newcommand*{\rest}{\!\!\upharpoonright} 

\makeatletter
\newcommand{\extp}{\@ifnextchar^\@extp{\@extp^{\,}}}
\def\@extp^#1{\mathop{\bigwedge\nolimits^{\!#1}}}
\makeatother

\usepackage[
maxbibnames=10,
backend=biber,
giveninits=true, doi=false, isbn=false, url=false , 
mcite = true,
style=ieee-alphabetic
]{biblatex}
\renewbibmacro{in:}{%
	\ifentrytype{article}{\printtext{$\,$}}{\printtext{\bibstring{in}\intitlepunct}}}
\DeclareFieldFormat[article]{citetitle}{#1}   
\DeclareFieldFormat[article]{title}{#1} 
\AtEveryBibitem{\clearfield{month}}
\AtEveryBibitem{\clearfield{day}}
\AtEveryBibitem{\ifentrytype{book}{\clearfield{pages}}{}} 
\AtEveryBibitem{\clearfield{pagetotal}}  
\DeclareFieldFormat[book,inbook,incollection,inproceedings]{series}
{#1}
\bibliography{biblio.bib}  

\title{Review and concrete description of the irreducible unitary representations
  of the universal cover of the complexified Poincar\'e group}
\author{L. M. Borasi}

\author{Luigi M. Borasi\\[.5em]
  \tt{\small borasi@iam.uni-bonn.de}\\
  \tt{\small luigi.borasi@gmail.com}
}

\begin{document}
\maketitle

\begin{abstract} 
  We give a pedagogical presentation of the irreducible unitary representations of $\CC^4\rtimes\Spin(4,\CC)$,
  that is, of the universal cover of the complexified Poincar\'e group $\CC^4\rtimes\SO(4,\CC)$.
  These representations were first investigated by Roffman in 1967.
  We provide a modern formulation of his results
  together with some facts from the general Wigner-Mackey theory which are relevant in this context.
  Moreover, we discuss different ways to realize these representations and, in the case of a non-zero ``complex mass'',
  we give a detailed construction of a more explicit realization.
  This explicit realization parallels and extends 
  the one used in the classical Wigner case of $\RR^4\rtimes\Spin^0(1,3)$.
  Our analysis is motivated by the interest in the Euclidean formulation of Fermionic theories.
\end{abstract}
\section{Introduction}\label{Intro}
Wigner~\cite{wigner_unitary_1939} gave a complete description of the irreducible unitary representations of
$\RR^4\rtimes\Spin^0(1,3)$
and, roughly speaking, reinterpreted these representations, physically, as elementary particles
(for a list of the original contributions we refer to e.g~\cite[p.\ 30]{streater_pct_1978}
and~\cite[p.\ 105]{weinberg_quantum_1995}).
In the axiomatic formulation of quantum field theory~\cite{streater_pct_1978},
in particular in the context of the Wightman functions,
it is natural to extend the group $\RR^4\rtimes\Spin^0(1,3)$
to its complexification $\CC^4\rtimes\Spin(4,\CC)$.
Therefore, it makes sense to shift the attention from  $\RR^4\rtimes\Spin^0(1,3)$
to its complexification $\CC^4\rtimes\Spin(4,\CC)$
and study the representation theory of the latter
by an appropriate extension of the classical Wigner analysis~\cite{wigner_unitary_1939}.
To the best knowledge of the author, this complexified case was first studied 
by Roffman~\cite{roffmanUnitaryNonunitaryRepresentations1967}
(cf.\ also~\cite{roffman_complex_1966,roffmanGroupTheoreticApproachRelativistic1968,
  kawaiNonunitaryRepresentationsPoincare1969,beltramettiRepresentationsPoincareGroup1965})
who gave a complete description of the irreducible unitary representations arising in this case.

Our objective here is twofold.
On one hand, we want to give a self contained, modern presentation of the results in~\cite{roffmanUnitaryNonunitaryRepresentations1967}.
On the other hand, we want to describe in detail a way of realizing the irreducible unitary representations of
$\CC^4\rtimes\Spin(4,\CC)$
which is more directly related to the realization employed in the classical Wigner analysis.
In particular we describe in detail how the notion of ``standard boost'', employed in the classical Wigner analysis,
is generalized to the complexified case.

We distinguish two ways to realize these representation:
one which is ``implicit'' (Definition~\ref{def:induced_rep_0}),
in the sense that the representation is defined on a space of functions
satisfying a certain compatibility condition,
and another which is ``explicit'' (Corollary~\ref{cor:induced_rep_2}),
in the sense that the compatibility condition has been explicitly ``resolved''.
The implicit realization is more natural from the abstract point of view
and it is the one employed in~\cite{roffmanUnitaryNonunitaryRepresentations1967}
(compare~\cite[p.\ 250]{roffmanUnitaryNonunitaryRepresentations1967} and Definition~\ref{def:induced_rep_0} below).
The explicit one is useful in concrete applications
and gives a direct generalization of the realization employed by Wigner.
We will first describe this explicit realization in the general case of locally compact groups.
Then, after reviewing the results in~\cite{roffmanUnitaryNonunitaryRepresentations1967},
we apply this explicit realization to the irreducible unitary representations of $\CC^4\rtimes\Spin(4,\CC)$
corresponding to non-zero ``complex mass''.

To obtain the explicit realization,
we need to ``resolve'' the compatibility condition of the implicit realization.
This is done by choosing a Mackey decomposition for $\Spin(4,\CC)$,
that is, a measurable section $\Spin(4,\CC)/\mathbf W\rightarrow \Spin(4,\CC)$,
where $\mathbf W$ denotes a little group (in the sense of Wigner).
It is more convenient, in place of defining a section,
to define the following map $\beta$ which, in the text below, we shall name a \textit{Wigner-Mackey embedding}.
We let  
$\beta:\mathscr O_{\mathring v}\rightarrow\Spin(4,\CC)$
be a measurable map satisfying some regularity conditions (cf.\ Subsection~\ref{subsec:concrete}),
and such that the following ``Mackey compatibility condition'' holds: $\beta(v)\cdot\mathring v= v$,
$v\in\mathscr O_{\mathring v}$,
where $\mathscr O_{\mathring v}$ denotes the $\Spin(4,\CC)$-orbit of a vector
$\mathring v \in\CC^4$ 
and
$\cdot$ denotes the action of $\Spin(4,\CC)$
on $\CC^4$.
As will be explained in Subsection~\ref{subsec:beta} a Wigner-Mackey embedding $\beta$
extends the notion of ``standard boost'' 
employed in the classical Wigner analysis.
 
By its very nature, the map $\beta$ is in general not unique.
To describe the situation in detail, we explicitly construct, in Subsection~\ref{subsec:beta},
three possible choices of such a Wigner-Mackey embedding.

We believe this analysis to be useful in future applications,
which we intend to discuss in a future publication,  in the context of the Euclidean Dirac Fields. 
Regarding this point cf.\ the Conclusion.

A brief account of the content is as follows.

In Section~2 we introduce basic definitions and some notation regarding
the spin groups, the universal cover of the Poincar\'e group, and its complexification.

In Section~3 we review the Wigner-Mackey theory of induced representations
applied to the representation theory of locally compact groups
which are ``regular semidirect products with Abelian normal subgroup''.
This is a classical topic, of which we try to give a concise modern review.
In Subsection~3.2 we define what we call a Wigner-Mackey embedding and use it
to obtain the natural generalization
of the realization of induced representation employed by Wigner.

In Section~4 and 5 we specialize to the case of the Lie group $\CC^4\rtimes\Spin(4,\CC)$.
In Section~4 we describe the decomposition of the orbits and the corresponding
little groups. Here we essentially reformulate the results in~\cite{roffmanUnitaryNonunitaryRepresentations1967}
and, for completeness, we give full proofs of the various results we cite.
These results suffice to characterize the irreducible unitary representations of $\CC^4\rtimes\Spin(4,\CC)$.
In Section~5 we specialize to the case of non-zero ``complex mass'' and
we describe how to concretely construct the irreducible unitary representations,
arising in this case,
on a space of square integrable functions supported on an orbit.
In Subsection~5.1 we construct and discuss three concrete ``Wigner-Mackey embeddings'' for the case of non-zero complex mass.   

We conclude with some closing remarks regarding our analysis.

\section{Universal cover of the complexified Poincar\'e group}\label{sec:ispin}
The purpose of this section is to define the universal cover of the complexified Poincar\'e group.
More details regarding the standard facts described in this section can be found
e.g.\ in~\cite{lawson_spin_1989,taylorNoncommutativeHarmonicAnalysis1986,fulton_representation_1991}.
Throughout this article, we shall follow the convention of calling Poincar\'e group the group $\RR^4\rtimes\SO^0(1,3)$
which more precisely should be named \textit{proper orthochronous Poincar\'e group}.
In the first subsection we first introduce some notation regarding Clifford algebras and then
we introduce the Spin groups and the inhomogeneous Spin groups.
In the second subsection we discuss a standard representation of the complex Clifford algebra
over $\CC^4$ which is convenient in applications.
We shall use this representation in Sections~\ref{sec:wigner} and~\ref{sec:non-zero-mass}.

\subsection{$\Spin(4,\CC)$ and  $\ISpin(4,\CC)$}
Consider a field $\KK=\RR \text{ or } \CC$ and a vector space $V$ over $\KK$.
Let $B:V\times V\rightarrow\KK$ be a non-degenerate symmetric bilinear form on $V$.
We denote by $\Cliff(V,B,\KK)$ the Clifford algebra over $V$ obtained as the quotient of the
tensor algebra $\mathcal TV$ by the ideal generated by elements of the form $v\otimes v - B(v,v)$.
Notice the minus sign.
This convention is standard in the physical literature but is opposite e.g.\ to the one taken in~\cite{lawson_spin_1989}.
Let $V_\RR$ be a real vector space with a non-degenerate symmetric bilinear form $B_\RR$.
Let $W_\CC\deq\CC\otimes_\RR V_\RR$ be the complexification of $V_\RR$ and $B_\CC$
the complex-bilinear extension of $B_\RR$ to $W_\CC$.
Then the complex Clifford algebra $\Cliff(W_\CC,B_\CC,\CC)$ is isomorphic to the complexification
$\CC\otimes_\RR\Cliff(V_\RR,B_\RR,\RR)$ (cf.~\cite[p.\ 27]{lawson_spin_1989}).
We remark that we have a natural embedding of $V$ into the tensor algebra $\mathcal TV$ simply
by sending $V$ into the copy of $V$ in the tensor algebra.
This embedding descends to a natural embedding of $V$ into $\Cliff(V,B,\KK)$.

We now specialize to the concrete cases we will need in the rest of the article.
Let $\eta$ be the Minkowski non-degenerate symmetric bilinear form on $\RR^4$
with signature $(1,3)$. Without loss of generality we can assume
$\eta=\text{diag}(1,-1,-1,-1)$.
Moreover, let $\delta$ denote the standard Euclidean bilinear form on $\RR^4$.
We shall consider $\eta$ and $\delta$ also as bilinear forms on $\CC^4$
without introducing additional notation.

We denote by $\Cliff(1,3)\deq\Cliff(\RR^4,\eta,\CC)$ the real Clifford algebra on $\RR^4$
with respect to $\eta$ and $\CCliff(4)\deq\Cliff(\CC^4,\delta,\CC)$ the complex Clifford algebra over $\CC^4$.
We remark that on $\CC^n$, $n\in\NN$, all non-degenerate
symmetric bilinear forms are equivalent,
therefore $\CCliff(4)$ is isomorphic to the complexification
$\Cliff(\CC^4,\eta,\CC)$ of $\Cliff(\RR^4,\eta,\RR)$.
Hence, when considering $\CCliff(4)$, we are in a sense treating both the Euclidean and Minkowski case at once.
Note that, even though $\CCliff(4)=\Cliff(\CC^4,\delta,\CC)\cong\Cliff(\CC^4,\eta,\CC)$,
the natural embedding of $\CC^4$ into $\CCliff(4)$
is different from the natural embedding of $\CC^4$ into $\Cliff(\CC^4,\eta,\CC)$.
To encode this point in our notation we introduce,
alongside the natural embedding of $\CC^4$ into $\CCliff(4)$,
the following embedding which will make
the relation between $\CCliff(4)$ and $\Cliff(1,3)$ more transparent.

We let $\gamma:\CC^4\rightarrow\CCliff(4)$ be the embedding defined
by first sending an element $v=(v_0,v_1,v_2,v_3)\in\CC^4$ to
$\tilde v\deq(v_0,\ii v_1,\ii v_2,\ii v_3)$ and then embedding $\tilde v$
into $\CCliff(4)$ by the natural embedding of $\CC^4$ into $\CCliff(4)$.
We can express this in a basis independent way by defining
$\gamma:\CC^4\rightarrow\CCliff(4)$ to be the embedding
which satisfies the following relation
\begin{equation}\label{eq:Minkowski embedding}
  \gamma(v) = \PP_0v + \ii(\id-\PP_0)v,\quad v\in\CC^4
  ,
\end{equation}
where the left hand side is considered as en element of $\CCliff(4)$
(via the natural embedding of $\CC^4$ into $\CCliff(4)$),
and $\PP_0$ denotes the orthogonal projection onto the one-dimensional
space corresponding to the $+1$ eigenvalue of $\eta$.
For future convenience, let us name the embedding $\gamma:\CC^4\rightarrow\CCliff(4)$ 
the \textbf{Minkowski embedding} of $\CC^4$ into $\CCliff(4)$.
Note that we have
\begin{equation}
  \label{eq:gamma_anticommutators}
  \{\gamma(v),\gamma(w)\}=2\eta(v\cdot w)\id,
  \quad v,w\in\CC^4
  ,
\end{equation}
where $\{A,B\}\deq AB+BA$ denotes the anticommutator (for $A,B\in\CCliff(4)$) and $\id$
denotes the identity of $\CCliff(4)$.
If we call an element $v=(v_0,v_1,v_2,v_3)\in\CC^4$ real when
its components $v_\mu$ are real, $\mu\in\{0,1,2,3\}$,
then we see from~\eqref{eq:gamma_anticommutators}
that the image of $\gamma$, restricted to real elements,
generates a real subalgebra of $\CCliff(4)$ isomorphic to $\Cliff(1,3)$.

Let $\Spin(4,\CC)\subset\CCliff(4)$ be the Spin group over $\CC^4$
defined as follows (cf.\ e.g.~\cite[Section 20.2, formula (20.27), p.\ 308]{fulton_representation_1991}):
\begin{multline}\label{eq:SpinC}
  \Spin(4,\CC) \deq \big\{  v_1\cdots v_{2r} \in\CCliff(4) \st\\
  v_j\cdot v_j=\pm 1,\, v_j\in\CC^4,\,
  j=1,\dots,2r,\, r\in\NN 
  \big\}
  .
\end{multline}
The group $\Spin(4,\CC)$ has, canonically, both the structure of an \textit{analytic group} (of dimension $6$),
and the structure of a (real) \textit{Lie group} (of dimension $12$)\footnote{
  We can trivially consider $\CC^n$ as a \textit{real} vector space simply by allowing
  the scalars to be only reals.
  In a similar way any complex manifold can be looked upon as a real manifold.
  Moreover, any analytic group can be looked upon as a (real) Lie group.
  A basis $\mathfrak B$ of a complex vector space $V$
  induces a basis $\mathfrak B_\RR \deq \{ v\st v\in\mathfrak B\}\cup\{\ii v \st v\in\mathfrak B\}$ of $V$ as a real vector space.
  For example if
  $\mathfrak B \deq \big\{ a_{\mu\nu} = \tfrac12 v_\mu v_\nu\st v_\mu,v_\nu\in\CC^4\rightarrow\CCliff(4), \mu,\nu\in\{0,1,2,3\}\big\}$
  is a basis  for the \textit{complex} Lie algebra $\mathfrak{so}(4,\CC)$,
  then $\big\{ a \st a \in \mathfrak B\big\} \cup \big\{ \ii a \st a \in \mathfrak B\big\}$
  is a basis of $\mathfrak{so}(4,\CC)$ considered as a \textit{real} Lie algebra.
  In particular the element $a\in\mathfrak{so}(4,\CC)$ and the element $\ii a$
  are linearly independent when $\mathfrak{so}(4,\CC)$ is considered as a real Lie algebra.  
}.
Since we are interested in its unitary representations, unless we specify otherwise, we will consider it as a (real) Lie group.
In terms of the Minkowski embedding $\gamma$ defined above, we have
\begin{multline}\label{eq:SpinC}
  \Spin(4,\CC) =\big\{  \gamma(v_1')\cdots \gamma(v_{2r}') \in\CCliff(4) \st\\
  \eta(v_j',v_j')=\pm 1,\, v_j'\in\CC^4,\,
  j=1,\dots,2r,\, r\in\NN
  \big\}
  .
\end{multline}
Let $e_\mu$, $\mu\in\{0,1,2,3\}$, be the standard basis of $\CC^4$.
We define the following ``volume form'' on $\CCliff(4)$:
\begin{equation}
  \label{eq:volume_form}
  \Omega\deq e_0e_1e_2e_3\in\CCliff(4)
  ,
\end{equation}
where we denoted by juxtaposition the multiplication in $\CCliff(4)$.
In terms of the embedding $\gamma$ we have
\begin{equation}
  \label{eq:volume_form_gamma}
  \Omega = -\ii\gamma(e_0)\gamma(e_1)\gamma(e_2)\gamma(e_3)
  .
\end{equation}
What we denote here $\gamma(e_\mu)$, $\mu\in\{0,1,2,3\}$,
are the standard gamma matrices $\gamma_\mu$
in the physical literature considered as abstract objects without
picking a representation (e.g. Dirac, Weyl, Majorana, etc\dots).
In the next subsection we will review the Weyl representation.
In the standard physical notation our $\Omega$ is just $\Omega = -\gamma_5$
where $\gamma_5\deq\ii\gamma_0\gamma_1\gamma_2\gamma_3$,
$\gamma_\mu$, $\mu\in\{0,1,2,3\}$,
are the standard gamma matrices which satisfy:
$\{\gamma_5,\gamma_\mu\}=0$, $\gamma_5^2=\id$, $\gamma_5^\ast=\gamma_5$,
$\{\gamma_\mu,\gamma_\nu\} = 2\eta_{\mu\nu}$, $\eta_{00}=1$, $\eta_{jj}=-1$, $j=1,2,3$, $\mu,\nu\in\{0,1,2,3\}$.

Employing the canonical embedding $\CC^4\rightarrow\CCliff(4)$, we define the following canonical action $\lambda$
of $\Spin(4,\CC)$ on $\CC^4$:
\begin{equation}
  \label{eq: action}
  \lambda({s})v \deq svs^{-1},
  \quad v\in\CC^4\subset\CCliff(4),\quad {s}\in\Spin(4,\CC)
  .
\end{equation}
It is a standard result
(cf.\ e.g.~\cite[Chapter 12, (1.43) p.\ 251]{taylorNoncommutativeHarmonicAnalysis1986})
that the action $\lambda$ induces the covering map
\begin{equation*}
  \Spin(4,\CC) \rightarrow \SO(4,\CC)
  .
\end{equation*}

We now define\footnote{
  The notation $\ISpin$  is not standard in the literature.
  The symbol $\ISpin$ is meant as shorthand for \textit{inhomogeneous Spin group}
  In employing this terminology we mimic a similar convention
  for the Euclidean group in $n$ dimensions
  which is sometime denoted $\textbf{ISO}$
  for \textit{inhomogeneous special orthogonal group}.}
\begin{equation*}\label{eq:ISpin}
  \ISpin(4,\CC)\deq \CC^4\rtimes_\lambda \Spin(4,\CC)
  ,
\end{equation*}
where $\rtimes_\lambda$ denotes the semidirect product with respect to
the action $\lambda$.
If we denote by $(z,s)$ an element of $\ISpin(4,\CC)$, $z\in\CC^4$, $s\in\Spin(4,\CC)$,
then we have the following product rule which defines the semidirect product $\rtimes_\lambda$:
\begin{equation*}
  (z,s)(z',s') = (z+\lambda(s)z',ss'),
  \quad z,z'\in\Spin(4,\CC), \quad z,z'\in\CC^4.
\end{equation*}
For readability sake, we shall often denote $\lambda(s)v$, $s\in\Spin(4,\CC)$, $v\in\CC^4$,
simply by $s\cdot v$ and we shall write simply
$\CC^4\rtimes \Spin(4,\CC)$ where the reference to the action $\lambda$ is dropped.

We also consider the Spin group over $\RR^4$ with respect to $\eta$,
which has the following definition
(cf.\ e.g.~\cite[Chapter 12, formula (1.42), p.\ 251]{taylorNoncommutativeHarmonicAnalysis1986}):
\begin{multline*}
  \Spin(1,3) \deq \big\{ x_1\cdots x_{2r} \in\Cliff(1,3) \st \\
  \eta(x_j, x_j)=\pm 1,\, x_j\in\RR^4,\,
  j=1,\dots,2r,\, r\in\NN 
  \big\}
  .
\end{multline*}
We denote by $\Spin^0(1,3)$ the component of $\Spin(1,3)$ connected with the identity.
It is a standard fact that $\Spin^0(1,3)$ is isomorphic to $\SL(2,\CC)$ considered as a real Lie group.
We will describe the isomorphism concretely in the following subsection where we introduce a concrete representation.
Moreover, let us define
$$
\ISpin^0(1,3)\deq\RR^4\rtimes_\lambda\Spin^0(1,3),
$$
where $\lambda$ is the action defined in~\eqref{eq: action}, now restricted to $\Spin^0(1,3)$.
It is well known that $\ISpin^0(1,3)$ is the universal cover of the proper orthochronous Poincar\'e group
$\mathbf{ISO}^0(1,3)\deq\RR^4\rtimes\SO^0(1,3)$ 
(cf.\ e.g.~\cite[Chapter 12, section 1]{taylorNoncommutativeHarmonicAnalysis1986}).

We conclude this section noting the following standard facts.
The complex Spin groups $\Spin(n,\CC)$, for $n\ge 3$,
is connected, simply-connected,
and it is isomorphic to the universal cover of the group $\SO(n,\CC)$
(cf.\ e.g~\cite[Proposition 20.28, p.\ 308]{fulton_representation_1991} for complex case (as we have here)
and~\cite[Theorem 2.10, p.\ 20]{lawson_spin_1989} for the analogous real case).
From this it follows at once that the group
$\ISpin(4,\CC)$ is isomorphic to the universal cover of the complexified Poincar\'e group
$\mathbf{ISO}(4,\CC)\deq\CC^4\rtimes\SO(4,\CC)$.

\subsection{Weyl representation}\label{par: Weyl basis}
It is often convenient to represent the Clifford algebra $\CCliff(4)$
as the algebra $\mathrm M(4,\CC)$ of complex $4\times4$ matrices.
One such representation is given by Weyl's choice of matrix generators of $\CCliff(4)$.
Let $e_\mu$, $\mu=0,1,2,3$, be the standard basis of $\CC^4$  
and consider the Minkowski  embedding $\gamma:\CC^4\rightarrow\CCliff(4)$ defined in~\eqref{eq:Minkowski embedding}.

Then we define the \textbf{Weyl representation}
$\gamma^{\textrm{Weyl}}:\CCliff(4)\rightarrow \mathrm M(4,\CC)$
of $\CCliff(4)$
by associating, for $\mu\in\{0,1,2,3\}$,
$\gamma(e_\mu)\in\CCliff(4)$ to $\gamma^{\textrm{Weyl}}(e_\mu) \deq \gamma_\mu^{\textrm{Weyl}}\in\mathrm M(4,\CC)$, where
\begin{equation*}
  \gamma_0^{\textrm{Weyl}}\deq
  \begin{pmatrix}
    0&\id_2\\
    \id_2&0
  \end{pmatrix},
  \quad
  \gamma_j^{\textrm{Weyl}}\deq
  \begin{pmatrix}
    0&\sigma_j\\
    -\sigma_j&0
  \end{pmatrix},
  \;
  j=1,2,3
  ,
\end{equation*}
with $\id_2$ the unit $2$-by-$2$ matrix, and
$\sigma_1=\begin{psmallmatrix}0&-\ii\\\ii&0\end{psmallmatrix}$,
$\sigma_2=\begin{psmallmatrix}0&1\\1&0\end{psmallmatrix}$,
and $\sigma_3=\begin{psmallmatrix}1&0\\0&-1\end{psmallmatrix}$
the standard Pauli matrices.

The representation $\gamma^{\textrm{Weyl}}$ of $\CCliff(4,\CC)$
naturally restricts to a representation of $\Spin(4,\CC)$.
Given an element $s\in\Spin(4,\CC)$ in the image of the exponential map,
its representation $\gamma^{\textrm{Weyl}}(s)$
is explicitly given as follows, for some $w_{\mu\nu}\in\CC$,
$0\le\mu<\nu\le3$,
\begin{equation*}
  \gamma^{\textrm{Weyl}}(s)
  = \exp\Big\{
  \sum_{0\le \mu<\nu\le 3} w_{\mu\nu}
  \tfrac{1}{2}\gamma^{\textrm{Weyl}}_\mu\gamma^{\textrm{Weyl}}_\nu \Big\}
  .
\end{equation*}
By explicitly evaluating the generators $\tfrac{1}{2}\gamma_\mu\gamma_\nu$
of the Lie algebra $\mathfrak{so}(4,\CC)$ of $\Spin(4,\CC)$,
one obtains that any element $s\in\Spin(4,\CC)$ is represented by a matrix of the form
\begin{equation*}
  \begin{pmatrix} A&0\\0&B\end{pmatrix}
  ,
\end{equation*}
for some $A,B\in\SL(2,\CC)$.
This representation puts in evidence the following standard Lie group isomorphism:
\begin{equation*}
  \Spin(4,\CC) \cong \SL(2,\CC)\times\SL(2,\CC)
  .
\end{equation*}

Consider now the action $\lambda$ defined in~\eqref{eq: action}.
In view of the isomorphism
$\Spin(4,\CC)\cong\SL(2,\CC)\times\SL(2,\CC)$,
a convenient and standard representation of this action
is obtained as follows.
Let us identify $\CC^4$ with the vector space $\mathrm M(2,\CC)$
of $2$-by-$2$ complex matrices via the isomorphism
$\sigma:\CC^4\rightarrow \mathrm M(2,\CC)$ given by
\begin{equation}\label{eq:map sigma}
  v \mapsto \sigma(v)\deq {[v]}_0\id_2 + \sum_{j=1}^3 {[v]}_j\sigma_j
  ,
\end{equation}
where ${[v]}_\mu$, $\mu\in\{0,1,2,3\}$, denotes the $\mu$-th component
of $v\in\CC^4$ with respect to the standard basis of $\CC^4$,
$\id_2$ denotes the identity $2$-by-$2$ matrix,
and $\sigma_1,\sigma_2,\sigma_3$ are the standard Pauli matrices.
Now, because of the isomorphism $\Spin(4,\CC)\cong\SL(2,\CC)\times\SL(2,\CC)$,
we identify $s\in\Spin(4,\CC)$ with a pair $(A,B)\in\SL(2,\CC)\times\SL(2,\CC)$
where $A,B\in\SL(2,\CC)$. With this identification,
and the identification of $\CC^4$ with $\mathrm M(2,\CC)$,
we see that the action $\lambda$ in~\eqref{eq: action}
induces an action $\lambda'$ of $\SL(2,\CC)\times\SL(2,\CC)\cong\Spin(4,\CC)$
on $\mathrm M(2,\CC)\cong\CC^4$ given by
\begin{equation} 
  \label{eq: action 2}
  \lambda'_{(A,B)}(M) \deq AMB^{-1},
  \quad A,B\in\SL(2,\CC), M\in\mathrm M(2,\CC)
  .
\end{equation}
This shows that we can identify $\ISpin(4,\CC)$
with
$\mathrm M(2,\CC)\rtimes_{\lambda'}(\SL(2,\CC)\times\SL(2,\CC))$,
where the action of $\SL(2,\CC)\times\SL(2,\CC)$ on $\mathrm M(2,\CC)$
is given by~\eqref{eq: action 2}.

\section{Irreducible unitary representations of regular semidirect products with Abelian normal subgroup}\label{sec:rep}
For a locally compact group $\Gg$ which is a regular semidirect product (in the sense defined below)
of an Abelian locally compact group with another locally compact group,
the theory of induced representations
yields the complete characterization of the irreducible unitary representations
of $\Gg$. In fact, in this situation, every irreducible unitary representation of $\Gg$ is 
induced from a irreducible unitary representation of a subgroup.
We will state this standard result as Theorem~\ref{th:general classification}, after we have given the necessary definitions.
The general theory can be approached within different frameworks and with varying degree of generality
~\cite{barut_theory_1986, kaniuthInducedRepresentationsLocally2013,
  rieffelInducedRepresentationsAlgebras1972,
  fell_representations_1988-1,
  manuilovHilbertModules2005,
  raeburnMoritaEquivalenceContinuousTrace1998}.
We mainly follow~\cite{kaniuthInducedRepresentationsLocally2013}.

Loosely speaking, inducing a representation
is a natural way to extend a representation of a subgroup $H$ of a group $G$ to
a representation of $G$.
Since representations are considered only up to equivalence,
there exist different equivalent realizations of an induced representation
which can be taken as a definition.
In Subsection~\ref{subsec:general-theory}, we take as definition of an induced representation
perhaps the most standard one. This definition can be stated efficiently in a self-contained manner.
In Subsection~\ref{subsec:concrete} (cf. Corollary~\ref{cor:induced_rep_2}),
we describe a more explicit and concrete realization of an induced representation
which is, in a sense, the natural generalization
of the realization employed in the classical Wigner analysis of $\ISpin^0(1,3)$.

In Section~\ref{sec:wigner}, we will specialize this general analysis to the case of $\Gg=\ISpin(4,\CC)$.
In Section~\ref{sec:non-zero-mass},
we will apply the realization of induced representations, given in Corollary~\ref{cor:induced_rep_2},
to explicitly describe the irreducible unitary representations of $\ISpin(4,\CC)$
corresponding to the special case of non-zero complex mass.

\subsection{General theory}\label{subsec:general-theory}
Let $G$ be a generic topological group. By a unitary representation of $G$ we mean a pair $(\mathcal H,\pi)$
where $\mathcal H$ is a separable complex Hilbert space and $\pi$ is a continuous homomorphism $\pi:G\rightarrow\mathbf U(\mathcal H)$,
where $\mathbf U(\mathcal H)$ denotes the space of unitary operators on $\mathcal H$ endowed with the weak operator topology.

We let $\Gg$ be a locally compact group, that is, a topological group, which is also a Hausdorff, locally compact space.
Moreover we let $\Hg$ be a closed proper subgroup of $\Gg$.
Consider the space $\Gg/\Hg=\{g\Hg\st g\in\Gg\}$ of left cosets with its natural quotient topology.
We denote the natural action of $\Gg$ on $\Gg/\Hg$ by left-multiplication, simply by juxtaposition:
$g\in\Gg$ acts on $x\in\Gg/\Hg$ by $g\mapsto gx$.        

We shall implicitly equip any topological space with the Borel
$\sigma$-algebra which is the $\sigma$-algebra generated by the open sets.
Moreover a measurable function between topological spaces, unless explicitly stated, will be understood to be Borel measurable.

We recall that a regular Borel measure $\mu$ on $\Gg/\Hg$ is said to be quasi-invariant when
$\mu\not\equiv0$ and, for every $g\in\Gg$, $\mu_g$ and $\mu$ are mutually absolutely continuous,
where, for a Borel set $B\subset\Gg/\Hg$, $\mu_g(B)\deq \mu(gB)$ denotes the measure translated by $g\in\Gg$.
On $\Gg/\Hg$ there always exists a quasi-invariant regular Borel measure such that the Radon-Nikodym derivative
\begin{equation}\label{eq:Radon-Nykodym}
  \varrho_g(x) \deq \dd\mu_{\Gg/\Hg}(g^{-1}x)/\dd\mu_{\Gg/\Hg}(x),
  \quad
  g\in\Gg, x\in\Gg/\Hg
  ,
\end{equation}
is a continuous function on $\Gg\times\Gg/\Hg$
and satisfies the identity
\begin{equation}\label{eq:identity}
  \varrho_{gg'}(x) = \varrho_{g'}(x)\varrho_g(g'x),
  \quad g,g'\in\Gg, x\in\Gg/\Hg
  ,
\end{equation}
(cf.\ e.g.~\cite[Theorem 1.18, p.\ 16]{kaniuthInducedRepresentationsLocally2013}).
We will implicitly assume a quasi-invariant measure $\mu$  on $\Gg/\Hg$ to be of this type,
i.e.\ a regular Borel measure with a continuous Radon-Nikodym derivative which satisfies the identity~\eqref{eq:identity} above.
\begin{parpar}[Definition (induced representation).]\label{def:induced_rep_0}
  Let $\Gg$ be a locally compact group, $\Hg$ a closed subgroup,
  and $(\mathfrak H,\rho)$ be a unitary representation of $\Hg$.
  We define the \textbf{induced unitary representation} $(\mathcal H^{\mathfrak H}, \text{Ind}_\Hg^\Gg(\rho))$ of $\Gg$,
  induced from the unitary representation $(\mathfrak H,\rho)$ of $\Hg$, to be 
  as follows.
  $\mathcal H^{\mathfrak H}$ is defined to be the complex Hilbert space of (equivalence classes of) measurable functions
  $f:\Gg\rightarrow \mathfrak H$ satisfying
  \begin{equation}%
    \label{eq:compatibility}
    f(gh) = \rho(h^{-1})f(g),
    \quad g\in\Gg, h\in\Hg
    ,
  \end{equation}
  and $\|f\|\deq(f,f)<+\infty$;
  here $(\cdot,\cdot)$ denotes the scalar product on $\mathcal H^{\mathfrak H}$ given by
  \begin{equation}\label{eq:scalar-prod}
    (f_1,f_2) \deq \int_{\Gg/\Hg}{(f_1(g),f_2(g))}_{\mathfrak H} \dd\mu([g])
    ,
  \end{equation}
  where ${(\cdot,\cdot)}_{\mathfrak H}$ denotes the scalar product on $\mathfrak H$
  and $[g]=g\Hg\in\Gg/\Hg$ denotes the left $\Hg$-coset corresponding to $g\in\Gg$.  
  Note that, by~\eqref{eq:compatibility}, the map $g\mapsto {(f_1(g),f_2(g))}_H$, $f_1,f_2\in H$, $g\in\Gg$,
  is constant on left $\Hg$-cosets,
  hence the scalar product on $\mathcal H^{\mathfrak H}$, defined in~\eqref{eq:scalar-prod}, is well defined.

  We define the continuous homomorphism $g\mapsto \text{Ind}_\Hg^\Gg(\rho)(g)$
  of $\Gg$ into $\mathbf U(\mathcal H)$ as follows
  \begin{equation*}
    \text{Ind}_\Hg^\Gg(\rho)(g)f(g_0) \deq \sqrt{\varrho_g([g_0])} f(g^{-1}g_0),
    \quad g,g_0\in\Gg,\quad f\in\mathcal H
    ,
  \end{equation*}
  where $\varrho_g([g_0])$, $g,g_0\in\Gg$, $[g_0]=g_0\Hg$, denotes the Radon-Nikodym derivative as in~\eqref{eq:Radon-Nykodym}. 
  The term $\sqrt{\varrho_g([g_0])}$ on the right hand side has the purpose of making this representation unitary,
  as can be easily checked 
  (cf.\ e.g.~\cite[Proposition 2.27]{kaniuthInducedRepresentationsLocally2013} or~\cite[Section 16.1, Lemma 2, p.\ 475 ]{barut_theory_1986}).
  Note also that the representation is indeed continuous (for details we refer to~\cite[Section 1.4]{kaniuthInducedRepresentationsLocally2013}).
\end{parpar}
Let us assume that there exists 
a closed, normal, Abelian subgroup  $\Ng$ of $\Gg$
(normal here is in the group-theoretic sense: $gn g^{-1}\in\mathbf N$, for all $n\in\mathbf N$, $g\in\mathbf G$).
Given a (locally compact) group $\Gg$ we denote by $\hat{\Gg}$ the space of all equivalence classes of
irreducible unitary representations of $\Gg$.
In the case of an Abelian (locally compact) group $\Ng$, an element $\chi\in\hat\Ng$ is a character for $\Ng$,
that is, a one dimensional unitary representation
$\chi:\Gg\rightarrow\mathbf U(1)$, where $\mathbf U(1)$ denotes the Lie group of complex numbers with modulus equal to one.

We make $\hat{\Ng}$ into a (locally compact) $\Gg$-space by letting $\Gg$ act on $\hat{\Ng}$
by conjugation (which we denote by a dot $\cdot$):
\begin{equation}
  \label{eq:G-space}
  (g\cdot\chi)(n)\deq \chi(g^{-1}ng), \quad n\in\Ng\subset\Gg, g\in\Gg
  .  
\end{equation}
Given a character $\chi\in\hat{\Ng}$,
we denote by $\Gg_\chi\deq\{g\in\Gg\st g\cdot\chi = \chi\}$ the \textbf{isotropy group} of $\chi$
and by $\Gg(\chi)\deq\{\chi'\in\hat\Ng\st\chi'=g\cdot\chi, g\in\Gg\}$
the \textbf{orbit} of $\chi$ under the action of $\Gg$

\begin{parpar}[Definition (Mackey compatible subgroup).]\label{def:Mackey comp}
  Let $\mathbf G$ be a locally compact group. A closed normal Abelian subgroup $\mathbf N$
  of $\mathbf G$ is called a \textbf{Mackey compatible} subgroup if
  \begin{enumerate}
  \item for each irreducible unitary representation $U\in\hat{\mathbf G}$,
    its restriction $U\rest_{\mathbf N}$ to $\Ng$ lives on a $\Gg$-orbit in $\hat{\Ng}$ and
  \item for each character $\chi\in\hat{\mathbf N}$,
    the map $\phi_\chi:\Gg/\Gg_\chi\rightarrow\Gg(\chi)$,
    $\phi_\chi:g\Gg_\chi \mapsto g\cdot\chi$, $g\in\Gg$,
    is a homeomorphism of $\mathbf G/\mathbf G_\chi$ with the orbit $\mathbf G(\chi)$.
  \end{enumerate}
\end{parpar}
We quote a sufficient condition for Mackey compatibility which we will apply in the case of $\ISpin(4,\CC)$.
We refer to~\cite[Remark 4.26, p.\ 159, and Proposition 4.6, p.\ 155]{kaniuthInducedRepresentationsLocally2013}
for a proof.
\begin{parpar-noproof}[Proposition.]\label{pr:condition for Mackey comp}
  Let, as above, $\Gg$ be a locally compact group with a closed normal Abelian subgroup $\Ng$.
  Consider the space
  $\mathscr X\deq\hat\Ng/\Gg\deq\{\Gg(\chi)\st\chi\in\hat\Ng\}$
  of $\Gg$-orbits in $\hat\Ng$, endowed with the quotient topology.
  Assume $\mathscr X$ to be to be almost Hausdorff, that is,
  every closed subset $C\subset\mathscr X$ contains a subset
  which is: relatively open and dense in $C$, and Hausdorff.
  Moreover assume $\Gg/\Ng$ to be $\sigma$-compact, that is,
  $\Gg/\Ng$ is the unions of countably many compact subspaces.
  Then $\Ng$ is Mackey compatible.
\end{parpar-noproof}

We now specialize to the situation where the group $\Gg$ splits into a semidirect product
$\Gg=\Ng\rtimes_\lambda \Sg$ of an Abelian subgroup $\Ng$ with another locally compact group $\Sg$
which acts on $\Ng$ via an action $\lambda$.
If we denote an element $g\in\Gg$ as a pair $g=(n,s)$, $n\in\Ng$, $s\in\Sg$,
we have the following product rule for elements $g=(n,s),g'=(n',s')\in\Gg$
$$
(n,s)(n',s') = (n+\lambda(s)n', ss'),\quad n\in\Ng, s\in\Sg
.
$$
We consider $\Ng$ and $\Sg$ naturally as subgroups of $\Gg$ via the identifications
$\Ng\cong(\Ng,\id)$, $\Sg\cong(0,\Sg)$, where $\id$, respectively $0$, denotes the identity
of $\Sg$, respectively $\Ng$.
Under this identification $\Ng$ is a closed normal Abelian subgroup of $\Gg$.
Note moreover that $\Sg$ acts on $\Ng$ by conjugation: for $s\in\Gg$, $n\in\Ng$,
we have that 
$$
(\lambda(s)n,e) = (0,s)(n,e)(0,s^{-1})
.
$$
This is compatible with the $\Gg$-space structure we have given to $\hat\Ng$ in~\eqref{eq:G-space} above.
We are therefore justified in simplifying the notation by writing
\begin{equation*}
  s\cdot n = \lambda(s)n, \quad s\in\Sg,n\in\Ng
  .
\end{equation*}
With this notation we can rewrite the action of $\Gg$ on $\hat\Ng$ in~\eqref{eq:G-space} as
\begin{equation*}
  (g\cdot\chi)(n) = \chi(g^{-1}\cdot n),
  \quad n\in\Ng, g\in\Gg
  .
\end{equation*}
We shall also simplify the semidirect product  $\rtimes_\lambda$ notation to just $\rtimes$,
hiding the reference to the chosen action $\lambda$.
Note that the orbit $\Sg(\chi)$ of $\chi\in\hat\Ng$ under $\Sg$ coincides with the orbit $\Gg(\chi)$
under the full group $\Gg$ because the restriction of the action of $\Ng$ on $\hat{\Ng}$ is trivial
(because $\Ng$ is Abelian).
Moreover note that the isotropy group $\Gg_\chi$, of a character $\chi\in\hat\Ng$,
decomposes as the semidirect product
\begin{equation*}
  \Gg_\chi = \Ng\rtimes\Sg_\chi,
\end{equation*}
where $\Sg_\chi$ is the isotropy group of $\chi$ under the action of $\Sg$ on $\hat{\Ng}$.
For later convenience, we reserve the term \textbf{little group}
to $\Sg_\chi$, and continue to call $\Gg_\chi$ the isotropy group of $\chi\in\hat\Ng$. 
\begin{parpar}[Definition (regular semidirect product with Abelian normal subgroup).]
  Let $\Gg=\Ng\rtimes\Sg$ be a semidirect product of a locally compact group $\Sg$
  acting on locally compact Abelian group $\Ng$ such that
  $\Ng$, as a normal Abelian subgroup of $\Gg$, is Mackey compatible
  in the sense of Definition~\ref{def:Mackey comp}.
  Then we call $\Gg=\Ng\rtimes\Sg$ a \textbf{regular the semidirect product with Abelian normal subgroup}.
\end{parpar}
We have the following characterization of the irreducible unitary representations of
the regular semidirect product
$\Ng \rtimes \Sg$ with Abelian normal subgroup
(cf. \cite[Theorem 4.28, p.\ 168]{kaniuthInducedRepresentationsLocally2013}).
\begin{parpar-noproof}[Theorem.]\label{th:general classification}
  Let $\Gg\deq \Ng \rtimes\Sg$
  be a regular semidirect product with Abelian normal subgroup.
  Let $\mathscr X=\hat{\Ng}/\Sg=\{\Sg(\chi)\st\chi\in\hat{\Ng}\}$
  be a section of the $\Sg$-orbits
  in $\hat{\mathbf N}$.
  Then the set $\hat{\Gg}$ of equivalent classes of irreducible unitary representations of $\Gg$ is given by
  \begin{equation*}
    \hat{\Gg} =
    \big\{  \text{Ind}_{\Ng\rtimes \mathbf S_{\chi}}^\Gg(\chi\times\rho) :
    \rho\in \widehat{\mathbf S_{\chi}}, \chi\in \mathscr X   \big\}
    ,
  \end{equation*}
  where $\widehat{\mathbf S_{\chi}}$ denotes the set of equivalent classes of
  irreducible unitary representations of the little group $\Sg_\chi$
  of the character $\chi\in\mathscr X$.
\end{parpar-noproof}

\subsection{``Explicit'' realization of induced representations}\label{subsec:concrete}
Let us denote by $q:\Gg\rightarrow\Gg/\Hg$ the natural quotient mapping which sends
each element $g\in\Gg$ into its left coset $q(g)=g\Hg\in\Gg/\Hg$.
Note that the quotient map $q:\Gg\rightarrow\Gg/\Hg$
is continuous and open by definition of quotient topology.

By a \textbf{measurable section} $\tau:\Gg/\Hg\rightarrow\Gg$ we mean a measurable map 
such that $q(\tau(x))=x$, for all $x\in\mathbf G/\Hg$.
We will say that a measurable section $\tau$ is \textbf{locally bounded}
when the set $\tau(K)$ has compact closure for any $K$ compact.

For convenience, we now restrict to the case where
$\Gg$ is a \textit{second-countable} locally compact group.
By second-countable we mean that there exists a countable basis for the topology of $\Gg$.
This restriction is definitely still general enough for the case $\Gg=\ISpin(4,\CC)$.
We shall discuss the general situation, where the second-countability assumption is dropped,
in the Remark~\ref{rem:general-section} below.

When $\Gg$ is second-countable, we have the following classical result
(due to Mackey~\cite[Lemma 1.1]{mackeyInducedRepresentationsLocally1952})
which guarantees the existence of a locally bounded Borel measurable section
(cf.\ the Lemma~\ref{cor:borel-section} below): 
\begin{parpar-noproof}[Lemma (Mackey decomposition~\cite{mackeyInducedRepresentationsLocally1952}).]\label{th:Mackey}
  Let $\mathbf G$ be a second-countable locally compact group and let
  $\mathbf H$ be a closed subgroup of $\mathbf S$.
  Then there exists a Borel set $\mathscr B\subset\mathbf G$
  such that
  \begin{enumerate}
  \item $\mathscr B$ intersects every left $\mathbf H$ coset in
    exactly one point and
  \item for each compact subset $K\subset \mathbf G$, $q^{-1}(q(K))\cap \mathscr B$
    has a compact closure
    ($q:\Gg\rightarrow\Gg/\Hg$ denotes the quotient mapping and $q^{-1}$
    its inverse image).\qedhere
  \end{enumerate}
\end{parpar-noproof}
\noindent
To facilitate the discussion below,
we give an equivalent statement of Lemma~\ref{th:Mackey}.
For brevity, we only prove that the statement of the Lemma~\ref{cor:borel-section} below
is equivalent to the statement of Lemma~\ref{th:Mackey} above.
We refer to the reference~\cite{mackeyInducedRepresentationsLocally1952} cited above
for a proof of the Lemma~\ref{th:Mackey}.
\begin{parpar}[Lemma (Mackey decomposition, equivalent version).]\label{cor:borel-section}
  Let $\mathbf G$, $\mathbf H$, $\mathscr B$, and $q$ be as in the lemma above.
  Then every element $g\in\mathbf G$ can be uniquely represented in the form
  \begin{equation*}
    g = b_g h_g ,\quad b_g\in \mathscr B, \quad h_g=b_g^{-1}g\in\mathbf H
    ,
  \end{equation*}
  and the map $\tau:\Gg/\Hg\rightarrow\Gg$ defined by
  \begin{equation*}
    \tau(g\Hg) \deq q^{-1}(g\Hg)\cap\mathscr B = g\Hg\cap\mathscr B=b_g,
    \quad g\Hg\in\Gg/\Hg, 
    ,
  \end{equation*}
  is a Borel measurable section and it is locally bounded.
  Vice versa, the image of a locally bounded measurable section $\tau:\Gg/\Hg\rightarrow\Gg$
  is a Borel measurable subset $\mathscr B$ of $\Gg$ with the properties $1.$ and $2.$ in the lemma above.
\end{parpar}
\begin{proof}
  The decomposition $g=b_g h_g$, $h_g\in\Hg$, $b_g\in\mathscr B$
  is a consequence of the fact that $\mathscr B$ intersects
  each left coset in exactly one point. Hence
  $b_g$ is unique and so is $h_g =b_g^{-1} g$.
  Note that, since $g\Hg=b_g\Hg$, then $h_g\Hg = b_g^{-1}g\Hg=\Hg$, hence $h_g$ is indeed in $\Hg$.
  
  Let $B\in\mathscr B$ be a Borel set in the Borel $\sigma$-algebra obtained
  by restricting the Borel $\sigma$-algebra of $\Gg$ to $\mathscr B\in\Gg$.
  We want to show that $\tau$ is a Borel map, that is, that $E \deq\tau^{-1}(B)$ is Borel.
  Since $\Gg$ is a second-countable, locally compact, Hausdorff space,
  it is a Polish space (cf.~\cite[Theorem (5.3), p.\ 29]{kechrisClassicalDescriptiveSet1995}) 
  and therefore it can be made into a complete metric space. 
  The same holds true for $\Gg/\Hg$.
  Hence we can apply Kuratowski's theorem
  (cf.~\cite[Theorem 3.9, p.\ 21]{parthasarathyProbabilityMeasuresMetric1967})
  which states that if $\phi$ is a measurable bijective map between second-countable
  complete metric spaces then it sends
  Borel sets into Borel sets.
  Since $q:\Gg\rightarrow\Gg/\Hg$
  restricted to $\mathscr B$ is continuous and bijective,
  by Kuratowski's theorem we obtain that $E=q(B)$ is indeed Borel.

  The fact that $\tau$ is locally bounded follows directly from the definition
  of $\tau$ and property (b) of $\mathscr B$ in Lemma~\ref{th:Mackey}.

  We now turn to the last assertion of Lemma~\ref{cor:borel-section}.
  Given a locally bounded measurable section $\tau:\Gg\rightarrow\Gg/\Hg$,
  let its image be denoted by $\mathscr B$.
  Since $\Gg/\Hg$ is second countable and locally compact, it is $\sigma$-compact,
  i.e.\ it is a countable union of compact subspaces.
  Since $\tau$ is locally bounded, its image is a countable union of relatively compact
  subsets of $\Gg$, hence it is in particular a Borel set.
  The fact that $\mathscr B$ intersects every left coset in exactly one point follows
  from the fact that $\tau$ is a section, i.e.\ $q(\tau(x))=x$, $x\in\Gg/\Hg$.
  Finally, since $\tau$ is assumed to be locally bounded, it follows that,
  for $K$ compact, $q^{-1}(K)\cap\mathscr B$ is relatively compact
  and the proof is complete.
\end{proof}
A measurable section $\tau$ as in Lemma~\ref{cor:borel-section}, allows us to give a
realization of an induced unitary representation, equivalent to the one in Definition~\ref{def:induced_rep_0},
but where the carrier space consists of square integrable functions on the quotient space $\Gg/\Hg$.
We phrase this as the following proposition, the proof of which is straightforward
(cf.~\cite[Section 2.4, p.\ 73]{kaniuthInducedRepresentationsLocally2013} or~\cite[Section 16.1, Lemma 1, p.\ 473]{barut_theory_1986}).
\begin{parpar-noproof}[Proposition.]\label{pro:induced_rep_1}
  Let $\Gg$ be a locally compact group, $\Hg$ a closed subgroup,
  and $(\mathfrak H,\rho)$ be a unitary representation of $\Hg$.
  Assume there exists a locally bounded measurable section $\tau:\Gg/\Hg\rightarrow\Gg$.
  Then, the induced representation $(\mathcal H^{\mathfrak H},\text{Ind}_\Hg^\Gg(\rho))$
  of Definition~\ref{def:induced_rep_0}
  is unitarily equivalent to the representation $(\tilde{\mathscr H}, \tilde U_\Hg^\Gg)$
  where
  \begin{equation*}
    \tilde{\mathscr H} \deq L^2(\Gg/\Hg,\mu; \mathfrak H)
    ,
  \end{equation*}
  and, for all $g\in\Gg$, $x\in\Gg/\Hg$, and $F\in\tilde{\mathscr H}$,
  \begin{equation}\label{eq:action}
    \big(\tilde U_{\Hg}^{\Gg}(\rho)(g) F\big)(x)
    \deq  \sqrt{\varrho_g(x)} \,
    \rho\big({\tau(x)}^{-1} g \, \tau(g^{-1} x)\big)
    \big(F(g^{-1}x)\big)
    .
  \end{equation}
  Let $W:\mathcal H^H\rightarrow\tilde{\mathscr H}$, be the map
  \begin{equation*}
    (Wf)(x)\deq f(\tau(x)),
    \quad x\in\Gg/\Hg, f\in\mathcal H^H
    .
  \end{equation*}
  Then $W$ is an isometric isomorphism
  which sends $(\mathcal H^H,\text{Ind}_\Hg^\Gg(\rho))$
  into $(\tilde{\mathscr H},\tilde U_\Hg^\Gg)$ and in particular we have, for all $g\in\Gg$,
  $\tilde U_\Hg^\Gg(\rho)(g) = W\,\text{Ind}_\Hg^\Gg(\rho)(g)W^{-1}$.
  Finally its inverse is given by
  \begin{equation*}
    (W^{-1}F)(g) = \rho(g^{-1}\tau(q(g)))F(q(g))=\rho(h_g^{-1})F(q(g)),
    \quad g\in\Gg, F\in\tilde{\mathscr H}
    ,
  \end{equation*}
  where $b_g=\tau(q(g))$, $h_g=b_g^{-1}g$ with $g=b_g h_g$, as in Lemma~\ref{cor:borel-section}.
\end{parpar-noproof}
\begin{parpar}[Remark.]\label{rm:argument}
  We first note that, even if the section $\tau$ is not necessarily continuous,
  the representation $(\tilde{\mathscr H},\tilde{U}_\Hg^\Gg(\rho))$ is in fact continuous
  since the representation $(\mathcal H^{\mathfrak H},\text{Ind}_\Hg^\Gg(\rho))$ is continuous and
  the isomorphism $W:\mathcal H^{\mathfrak H}\rightarrow\tilde{\mathscr H}$ is an isometry.
  Next, we show that the argument of $\rho$ on the right hand side of~\eqref{eq:action}
  is indeed an element of $\mathbf H$.
  In our notation of left action of $\Gg$ on $\Gg/\Hg$
  an element $g\in\Gg$ acts on $x\in\Gg/\Hg$ by $x\mapsto gx$.
  By definition of section we have
  \begin{equation*}
    x = \tau(x)\Hg,\quad x\in\Gg/\Hg.
  \end{equation*}
  Hence
  \begin{equation*}
    gx = g\tau(x)\Hg = q(g\tau(x)),\quad x\in\Gg/\Hg, g\in\Gg
    .
  \end{equation*}
  Then the argument of $\rho$ on the right hand side of~\eqref{eq:action} becomes
  \begin{align*}
    {\tau(x)}^{-1} g \, \tau(g^{-1}x)
    &= {\big({\tau(g^{-1} x)}^{-1} g^{-1}\tau(x)\big)}^{-1} \\
    &= {\big({\tau(q(g^{-1}\tau(x)))}^{-1} g^{-1}\tau(x)\big)}^{-1} \\
    &= {\big({\tau(q(g'))}^{-1} g'\big)}^{-1}
  \end{align*}
  for $g'=g^{-1}\tau(x)$.
  Now, from Lemma~\ref{cor:borel-section} we have
  that, for any $g'\in\Gg$, $\tau(q(g'))=b_{g'}$ and ${\tau(q(g'))}^{-1}g'= b_{g'}^{-1}g = h_{g'}\in\Hg$.
  Hence the argument of $\rho$ on the right hand side of~\eqref{eq:action} is an element of $\Hg$.
\end{parpar}
This realization of an induced representation has some advantages and some disadvantages compared to
the realization in~\ref{def:induced_rep_0}.
In~\ref{def:induced_rep_0} the carrier space is a Hilbert space of functions from the whole group $\Gg$ into $\mathfrak H$
constrained to satisfy the compatibility condition in~\eqref{eq:compatibility}.
The realization we give in~\ref{pro:induced_rep_1} has the advantage of eliminating this compatibility condition
and therefore reducing the domain from $\Gg$ to $\Gg/\Hg$.
This comes at the cost of introducing a non-canonical section $\tau:\Gg/\Hg\rightarrow\Gg$.
Hence, this definition of an induced representation is convenient especially when
such a section can be explicitly given. 

In the applications where we have a normal, Abelian, Mackey compatible subgroup $\Ng$ of $\Gg$,
it is convenient to slightly modify this realization of induced representation
and transform the Hilbert space on which it is realized into a Hilbert space
of functions on an orbit.
Indeed this is the realization employed in the classical Wigner analysis of $\ISpin^0(1,3)$.
We shall now describe this modification in detail.
Let us fix a character $\chi\in\hat\Ng$ and consider the closed subgroup
$\Hg$ to be the isotropy group $\Gg_\chi$.
Moreover let us rename $\tau_\chi=\tau$ in this case, since it depends on the choice of character $\chi$.
By point $2.$ of the Mackey compatibility condition~\ref{def:Mackey comp},
we have an isomorphism $\phi_\chi$ of the quotient $\Gg/\Gg_\chi$
with the orbit $\Gg(\chi)$ of $\chi$ under $\Gg$.
We can therefore transport the measurable section $\tau_\chi$
along $\phi_\chi^{-1}$ to obtain an embedding $\beta_\chi=\tau\circ\phi_\chi^{-1}$
of the orbit $\Gg\cdot\chi$ into $\Gg$.
Note that, by point ${2.}$ of~\ref{def:Mackey comp},
we have $\beta_\chi(v)\cdot\chi = v$, $v\in\Gg(\chi)$.
For later use, we make this observation into a definition.
\begin{parpar}[Definition.]\label{def:Mackey-embedding}
  Let $\Gg$ be a second-countable, locally compact group,
  $\Ng$ a Mackey compatible, closed normal Abelian subgroup,
  and $\chi\in\hat\Ng$ a character.
  We call a locally bounded, measurable map $\beta:\Gg(\chi)\rightarrow\Gg$,
  a \textbf{Wigner-Mackey embedding} when it satisfies
  \begin{equation}
    \label{eq:mackey-embedding}
    \beta_\chi(v)\cdot \chi = v,
    \quad v\in\Gg(\chi)
    ,
  \end{equation}
  that is, for any $v$ in the orbit $\Gg(\chi)\subset\hat\Ng$,
  $\beta_\chi(v)$ is an element of $\Gg$ which maps $\chi$ into
  $v$ under the action of $\Gg$ on $\hat\Ng$.
\end{parpar}
We shall give in subsection~\ref{subsec:beta} three examples of Wigner-Mackey embeddings
for the case $\Gg=\ISpin(4,\CC)$.

Given a measurable, locally bounded Wigner-Mackey embedding $\beta_\chi$,
then it is straightforward to see that $\beta_\chi\circ\phi_\chi$
defines a locally bounded measurable section,
where $\phi_\chi$ is, as above, the isomorphism in point ${2.}$ of the Mackey compatibility condition~\ref{def:Mackey comp}.
Hence this Wigner-Mackey embedding is equivalent, in the Mackey compatible case,
to choosing a locally bounded measurable section $\tau$ (cf.\ Proposition~\ref{pro:induced_rep_1}).  

The reason for introducing the notion of Wigner-Mackey embedding is that we can use it to realize the induced representation on a Hilbert space of functions
supported on an orbit.
This is arguably more concrete than the Hilbert space in the realization of Proposition~\ref{pro:induced_rep_1},
which is a Hilbert space of functions on the quotient space $\Gg/\Hg$.
We state this new (equivalent) realization as the following simple corollary to Proposition~\ref{pro:induced_rep_1}.
\begin{parpar}[Corollary.]\label{cor:induced_rep_2} 
  Let $\Gg=\Ng\rtimes\Sg$ be a regular semidirect product with Abelian normal subgroup. 
  Let $\chi\in\hat{\Ng}$ be a character and let
  $(\mathfrak H,\chi\otimes\rho_\chi)$
  be a unitary representation of $\Gg_\chi=\Ng\otimes\Sg_\chi$.
  Moreover let $\phi_\chi:\Gg/\Hg\rightarrow\Gg(\chi)$
  be a homeomorphism which satisfies the Mackey compatibility condition~\ref{def:Mackey comp}.
  Let $\mu_{\Gg(\chi)}$ be the measure on $\Gg(\chi)$
  obtained by pushing forward, along $\phi_\chi$, the measure $\mu$ on $\Gg/\Hg$.
  Then, the induced representation $(\mathcal H^{\mathfrak H},\text{Ind}_{\Gg_\chi}^{\Gg}(\chi\otimes\rho_\chi))$ is
  unitarily equivalent to the representation $(\mathscr H, U)$ given by
  \begin{equation}\label{eq:action orbit} 
    \begin{aligned}
      \mathscr H &\deq L^2(\Gg(\chi),\mu_{\Gg(\chi)};\mathfrak H), \\
      \big(U(g) F\big)(v) &\deq
      \sqrt{\vartheta_g(v)} \, 
      \rho\big(\beta(v)^{-1} g\beta(g\cdot v)\big) 
      \big(F(g^{-1} \cdot v)\big),
    \end{aligned} 
  \end{equation}  
  for $g\in\Gg$, $v\in\Gg(\chi)$, $F\in\mathscr H$,
  where
  $\vartheta_g(x)\deq\dd\mu_{\Gg(\chi)}(g\cdot v)/\dd\mu_{\Gg(\chi)}(v)$
  denotes the Radon-Nikodym derivative (cf.~\eqref{eq:Radon-Nykodym}).
\end{parpar}
\begin{proof}
  As in the discussion preceding Definition~\ref{def:Mackey-embedding},
  let $\phi_\chi:\Gg/\Hg\rightarrow\Gg(\chi)$ be an isomorphism which satisfies point ${2.}$ of the Mackey compatibility condition~\ref{def:Mackey comp}.
  Then $\phi_\chi$ induces a natural isometric isomorphism $\Phi_\chi:\tilde{\mathscr H}\rightarrow\mathscr H$,
  $\Phi_\chi:\tilde{f}\mapsto f$, $\Phi(\tilde{f})(v)= \tilde{f}(\phi_\chi^{-1}(v))$, $\tilde{f}\in\tilde{\mathscr H}$, $v\in\Gg(\chi)$.
  By applying $\Phi_\chi$ to the right hand side of~\eqref{eq:action} we obtain at once~\eqref{eq:action orbit}
  as soon as we note that, for any $g\in\Gg$, $v\in\Gg(\chi)$, we have
  (directly from the definition of the $\phi_\chi$ in the Mackey compatibility condition~\ref{def:Mackey comp})
  that
  \begin{equation*}
    \phi_\chi^{-1}(g^{-1}\cdot v) = g^{-1}x,
    \quad
    \phi_\chi^{-1}(\beta_\chi(v)^{-1}g\beta(g^{-1}\cdot v)) =   \tau_\chi(x)^{-1}g \tau_\chi(g^{-1}x)
    ,
  \end{equation*}
  where
  $\tau_\chi\equiv\beta\circ\phi_\chi^{-1}$,
  $x\deq\phi_\chi^{-1}(v)$, and, as before $gx$ denotes the element in $\Gg/\Hg$ obtained by left translating $x\in\Gg/\Hg$
  by $g\in\Gg$.
\end{proof}
\begin{parpar}[Remark: general locally compact case.]\label{rem:general-section}
  Above, we have assumed $\Gg$ to be a \textit{second-countable} locally compact group
  and discussed the existence of a locally bounded, Borel measurable section
  following~\cite{mackeyInducedRepresentationsLocally1952}.
  For $\Gg$ a general locally compact group, it was proved in~\cite{kehletCrossSectionsQuotient1984}
  the existence of a locally bounded \textit{Baire} measurable section.
  As a consequence,
  there exists a bijection $\Hg\times\Gg/\Hg\rightarrow \Gg$
  such that itself and its inverse are locally bounded Baire maps and preserve measurability of sets.
  Everything we have said in this subsection can be generalized
  to the case of $\Gg$ a general locally compact group
  if we replace Borel measures by Baire measures and Borel measurable sections by Baire measurable sections.
  Indeed Haar measures and quasi-invariant measures on the quotient $\Gg/\Hg$ are known
  to be completion regular
  (cf.~\cite{kehletCrossSectionsQuotient1984},
  a Borel measure $\mu$ is completion regular when~\cite[p.\ 230]{halmosMeasureTheory1950}:
  for every Borel $E$ set there exist two Baire sets $A,B$ such that
  $A\subset E\subset B$ and $\mu_0(B\setminus A)=0$, where $\mu_0$ is the unique Baire contraction measure
  associated to a Borel measure $\mu$).
  Hence these measures can be replaced by a corresponding Baire measure without
  changing the meaning of the spaces of square integrable functions.
\end{parpar}

\section{Wigner-Mackey analysis for $\ISpin(4,\CC)$}\label{sec:wigner}
We now want to apply the general theory of induced representations,
described in the section above,
to the case where
$$
\Gg = \ISpin(4,\CC) = \CC^4\rtimes\Spin(4,\CC)
\cong\mathrm M(2,\CC)\rtimes(\SL(2,\CC)\times\SL(2,\CC))
.
$$

In the notation of the previous section, this corresponds to taking
$\Ng = \CC^4$
and
$\Sg = \Spin(4,\CC)$.
As we remarked in Section~\ref{sec:ispin}, both $\CC^4$ and $\Spin(4,\CC)$
are considered here as (real) Lie groups.

Given $w=(w_0,w_1,w_2,w_3)\in\CC^4$,
we define a character $\hat{w}\in\hat\CC^4$ by
\begin{equation*}
  \hat{w}(v) \deq
  \exp\big\{ \ii\Re \{\eta(w, z)\} \big\}
  ,
\end{equation*}
where $z=(z_0,z_1,z_2,z_3)\in\CC^4$
and $\Re$ denotes the real part.
Any character $\chi\in\hat\CC^{4}$ is of this form for some $w\in\CC^4$.

As before, we denote by $g\cdot \chi$ the action of an element $g\in\ISpin(4,\CC)$
on an element $\chi\in\hat\CC^4$.
Explicitly we have
\begin{equation*}
  (s\cdot\chi)(z) = \chi(\lambda(s)^{-1}z) = \chi(s^{-1}\cdot z)
  ,
\end{equation*}
where $\lambda(s)^{-1}$ is by definition $\lambda(s^{-1})$
and, as before, we employ the same notation $s\cdot$
for the actions of $s\in\Sg$ on different spaces.

Consider the Minkowski embedding $\gamma:\CC^4\rightarrow\CCliff(4)$
defined in~\eqref{eq:Minkowski embedding}.
Then a character $\hat w\in\hat\CC^4\cong\CC^4$ can be written as
\begin{equation*}
  \hat w(v) = \exp\big\{ \ii \Re\big(  \tfrac14\tr[ \gamma(w)\gamma(v) ]  \big) \big\}
  ,
\end{equation*}
where $\tr$ is the trace on the Clifford algebra $\CCliff(4)$ i.e.\ the linear functional
$\tr:\CCliff(4)\rightarrow\CC$ which satisfies the properties:
$\tr(v)=\tr(-v)=0$, $v\in\CC^4\subset\CCliff(4)$,
$\tr(XY)=\tr(YX)$, $X,Y\in\CCliff(4)$,
and $\tr(\id)=4\id$.
The action of $\Spin(4,\CC)$
on $\hat\CC^4$ is then
\begin{equation}\label{eq:action_on_characters}
  s\cdot \hat w(v)
  = 
  \hat w(s^{-1}\cdot v)
  =
  \exp\big\{ \ii\Re\big(  \tr[ s\gamma(w) s^{-1}\gamma(v)]  \big) \big\}
  = \widehat{(s\cdot w)}(v)
  ,
\end{equation}
where $s\in\Spin(4,\CC)$ and
$\hat w\in\hat\CC^4$.
In the following we will often identify $\hat\CC^4$ with $\CC^4$ by identifying
$\hat w(v)=\exp\{\ii\Re\{\eta(w\cdot v)\}\}$ with $w$.
By~\eqref{eq:action_on_characters} we see that, when we identify $\hat\CC^4$ with $\CC^4$, the action of
$\Spin(4,\CC)$ on $\hat\CC^4$
is the same as the original action of $\Spin(4,\CC)$ on $\CC^4$.
Hence we are justified in denoting by $s\cdot$, $s\in\Spin(4,\CC)$, the action of $\Spin(4,\CC)$ on both $\CC^4$ and $\hat\CC^4$.

\subsection{Orbit structure and little groups}
We want to study the orbits in $\hat\Ng=\hat\CC^4$ 
under the action of $\Gg=\ISpin(4,\CC)$ or,
which is the same, under the action of $\Spin(4,\CC)$
(because $\mathbf N$, being Abelian, acts trivially on itself by conjugation).
Note that the orbits under the action $\lambda$ of $\Spin(4,\CC)$ on $\hat\CC^4$
coincide with the orbits under the natural action of
$\SO(4,\CC)$ on $\CC^4$.

It will be convenient to employ the isomorphism $\Spin(4,\CC)\cong\SL(2,\CC)\times\SL(2,\CC)$
detailed in Subsection~\ref{par: Weyl basis}.
If, as before, we denote by $\eta$ the Minkowski bilinear form on $\CC^4$,
then we have $\det(\sigma(v))=\eta(v,v)$, $v\in\CC^4$,
where $\sigma$ is the map in~\eqref{eq:map sigma}.
The action of $\SL(2,\CC)\times\SL(2,\CC)$ on $\sigma(v)\in\mathrm M(2,\CC)$
given in Subsection~\ref{par: Weyl basis} preserves such a quadratic form.
Identifying $\hat\CC^4$ with $\CC^4$ and $\CC^4$ with $\mathrm M(2,\CC)$,
we get that, for
$\sigma(\hat v)\in\mathrm M(2,\CC)$ and,
for every $(A,B)\in\SL(2,\CC)\times\SL(2,\CC)$
\begin{equation*}
  \det( (A,B)\cdot\sigma(\hat v) ) = \det( \sigma(\hat v))
  .
\end{equation*}
We have the following characterization of the orbits in $\hat\CC^4$ under $\Spin(4,\CC)$.
This parallels closely the result for the classical Poincar\'e group
(cf.\ e.g.~\cite[Chapter 17, section 2.B]{barut_theory_1986}).
\begin{parpar}[Lemma (Roffman~\cite{roffmanUnitaryNonunitaryRepresentations1967}).]\label{lem:orbit structure}
  Let us identify $\hat\CC^4$ with $\mathrm M(2,\CC)$.
  Then the $\ISpin(4,\CC)$-orbits in $\hat\CC^4$
  are in one-to-one correspondence with one of the following
  disjoint subsets of $\mathrm M(2,\CC)$:
  \begin{itemize}
  \item[]  $\mathscr O_0^0 \deq \{ 0\in\mathrm M(2,\CC) \}$;
  \item[] $\mathscr O_0 \deq \{ M\in\mathrm M(2,\CC) \st M\ne 0, \det M=0\}$;
  \item[] $\mathscr O_{z_m^2} \deq \{ M\in\mathrm M(2,\CC)\st \det M=z_m^2 \}$,
    $z_m^2\in\CC\setminus\{0\}$.
  \end{itemize}
\end{parpar}
\begin{proof}
  The element $0$ in $\mathrm M(2,\CC)$ remains fixed  under the action
  of $\ISpin(4,\CC)$. Hence one orbit is given by the singleton
  $\mathscr O_0^0 =\{ 0\in\mathrm M(2,\CC)\}$.

  Identifying $\hat\CC^4$ with $\mathrm M(2,\CC)$
  and $\Spin(4,\CC)$ with $\SL(2,\CC)\times\SL(2,\CC)$
  respectively, we see that the determinant of each element $M\in\mathrm M(2,\CC)$
  is an invariant of the action of $\ISpin(4,\CC)$.
  Hence we see that the sets
  $\mathscr O_0^0$, $\mathscr O_0$, and $\mathscr O_{z^2}$, $z^2\in\CC$,
  are invariant under the action of $\Spin(4,\CC)$.
  
  In the complement $\mathrm M(2,\CC) \setminus \{0\}$ of $\mathscr O_0^0$
  consider the set $\mathscr O_0$ of elements with determinant equal to zero.
  Let $N$ be any such element.
  We claim that the set $\mathscr O_0$ coincides with one orbit,
  that is, any $N$ can be brought into one another
  by a transformation in $\Spin(4,\CC)\cong\SL(2,\CC)\times\SL(2,\CC)$.
  It will suffice to show that any $N$
  can be brought into the element
  $\begin{psmallmatrix}1&0\\0&0 \end{psmallmatrix}\in\mathscr O_0$.

  First note that $N$ is a rank-one matrix and therefore there exist
  vectors $n_1,n_2\in\CC^2$ such that $N=n_1\otimes n_2$.
  Here we employ the standard identification of
  $n_1\otimes n_2$ with a linear map from $\CC^2$ into $\CC^2$.

  Let us denote by $e_1,e_2$ the standard basis of $\CC^2$
  and by $[n_j]_1,[n_j]_2$
  the components $n_j\in\CC^2$, $j=1,2$.
  We let $n_j^\perp \deq (-[n_j]_2, [n_j]_1)$, $j=1,2$,
  and define
  \begin{equation*}
    S(n_j) \deq \det(T(n_j))^{-1/2} T(n_j),
    \quad
    T(n_j) \deq (e_1\otimes n_j + e_2\otimes n_j^\perp)
    .
  \end{equation*}
  Then, $S(n_j)\in\SL(2,\CC)$ and, as a straightforward computation shows, we have
  $S(n_1)n_1 = S(n_2)n_2 = e_1$.
  
  Now consider the $\SL(2,\CC)\times\SL(2,\CC)$
  action on $N=n_1\otimes n_2$. For $(A,B)\in\SL(2,\CC)\times\SL(2,\CC)$
  we have $(A,B)\cdot N = (An_1)\otimes (B^{-1\mathtt{t}}n_2)$,
  where $\mathtt t$ denotes transposition.
  Let $A(n_1)=S(n_1)^{-1}$, $B(n_2)=S(n_2)^{\mathtt t}$.
  Then $(A(n_1),B(n_2))\in\SL(2,\CC)\times\SL(2,\CC)$
  and $(A(n_1),B(n_2))\cdot N = e_1\otimes e_1 =
  \begin{psmallmatrix}1&0\\0&0 \end{psmallmatrix}$ as we claimed.
  
  It remains to prove that each set
  $\mathscr O_{z_m^2} = \{ M\in\mathrm M(2,\CC)\st \det M = z_m^2 \}$,
  for $z_m^2\in\CC$, coincides with an orbit.
  
  Let $z_m$ be a square-root of $z_m^2$ and note that both
  $z_m\id_2$ and $-z_m\id_2$ are in $\mathscr O_{z_m^2}$
  (but \textit{not} $e^{\ii\theta}z_m\id_2$ for $e^{\ii\theta}\ne\pm1$).
  Indeed the element $(-\id_2,\id_2)\in\SL(2,\CC)\times\SL(2,\CC)$
  sends one into the other.
  Moreover any element $M\in\mathscr O_{z_m^2}$ is obtained
  from $z_m\id_2$ by the action of an element of $\SL(2,\CC)\times\SL(2,\CC)$.
  Indeed, since by hypothesis $z_m\ne0$, each matrix in $\mathscr O_{z_m^2}$
  is invertible. In fact, as a manifold, $\mathscr O_{z_m}$
  is just a non-trivially ``rescaled'' $\SL(2,\CC)$. 
  Let $M\in\mathscr O_{z_m}$.
  We chose in $\ISpin(4,\CC)\cong\SL(2,\CC)\times\SL(2,\CC)$
  an element of the form $(A,\id_2)$, with $A=z_mM^{-1}\in\SL(2,\CC)$.
  Then, by the action of $\Spin(4,\CC)$ we have
  $(A,\id_2)\cdot M = z_m\id_2$.
  This concludes the proof.
\end{proof}
\begin{parpar}[Corollary.]\label{cor:IspinMackeyComp}
  $\Ng=\CC^4$ is Mackey compatible (in the sense of Definition~\ref{def:Mackey comp})
  in $\Gg=\ISpin(4)$.
\end{parpar}
\begin{proof}
  By the Lemma~\ref{lem:orbit structure}, we have that the orbit space
  $\mathscr X=\hat\Ng/\Gg$ is isomorphic to the disjoint union
  \begin{equation*}
    \mathscr O_0^0  \sqcup \mathscr O_0 \sqcup \bigsqcup_{z_m^2\in\CC}\mathscr O_{z_m^2}
    .
  \end{equation*}
  That is, it is isomorphic to the disjoint union of $\CC$ and the two
  ``extra points'' $\mathscr O_0$ and $\mathscr O_0^0$.
  Hence it verifies the conditions of Proposition~\ref{pr:condition for Mackey comp}
  and the proof is complete. 
\end{proof}

\begin{parpar}[Lemma (Roffman \cite{roffmanUnitaryNonunitaryRepresentations1967}).]\label{lem:little group structure}
  Let $\latexchi_{z_m^2}\in\mathscr O_{z_m^2}$, for a fixed $z_m^2\in\CC$,
  and $\latexchi_0\in\mathscr O_0$
  be two characters.
  Then the little groups $\Sg_{\latexchi_{z_m^2}}$, $\Sg_{\latexchi_0}$ are respectively isomorphic to
  \begin{equation*}
    \Sg_{\latexchi_{z_m^2}} \cong \SL(2,\CC), \qquad\Sg_{\latexchi_0}\cong \CC^2\rtimes \Spin(2,\CC)
    .
  \end{equation*}
\end{parpar}
\begin{proof}
  As in Section~\ref{sec:rep} we identify $\hat\CC^4$ with $\mathrm M(2,\CC)$
  and $\Sg=\Spin(4,\CC)$ with $\SL(2,\CC)\times\SL(2,\CC)$.

  We consider first the case of $\latexchi_{z_m^2}\in\mathscr O_{z_m^2}$.
  Without loss of generality let $\latexchi_{z_m^2} \deq z_m\id_2$.
  Indeed, by the proof of Lemma~\ref{lem:orbit structure},
  $\latexchi_{z_m^2}$ is in the orbit $\mathscr O_{z_m^2}$.
  Let $(A,B)\in\SL(2,\CC)\times\SL(2,\CC)$, then $(A,B)$
  is in the isotropy group of $\latexchi_{z_m^2}$ if and only if
  $\latexchi_{z_m^2} = (A,B)\cdot \latexchi_{z_m^2} = A\latexchi_{z_m^2} B^{-1}$.
  Since $\latexchi_{z_m^2}$ is a multiple of the identity matrix, this condition implies
  $A=B$.
  Hence the isotropy group $\Sg_{\latexchi_{z_m^2}}$ is isomorphic to the diagonal subgroup
  $(\SL(2,\CC)\times\SL(2,\CC))^{\text{diag}}\deq\{(A,B)\in\SL(2,\CC)\times\SL(2,\CC)\st A=B\}$,
  of $\SL(2,\CC)\times\SL(2,\CC)$
  which is itself naturally isomorphic to $\SL(2,\CC)$.
  This concludes the proof of the first isomorphism stated in the lemma.

  We turn to the case $\latexchi_0\in\mathscr O_0$.
  Without loss of generality we can consider the case where
  $\latexchi_0 \deq \PP_1$ with $\PP_1\deq\begin{psmallmatrix}1&0\\0&0\end{psmallmatrix}\in\mathrm M(2,\CC)$ is the projection
  onto the vector $e_1=(1,0)\in\CC^2$.
  The condition for $(A,B)\in\SL(2,\CC)\times\SL(2,\CC)$ to be in the little group
  of $\latexchi_0$ is now
  $A\PP_1B^{-1} = \PP_1$.
  A straightforward computation shows that this condition implies that
  $A,B\in\SL(2,\CC)$ are of the form
  \begin{equation}\label{eq:AB}
    A= \begin{pmatrix}z_\times&z_1\\ 0&1/z_\times \end{pmatrix},
    \;
    B= \begin{pmatrix}z_\times&0\\ z_2&1/z_\times \end{pmatrix}
    ,
  \end{equation}
  for $z_\times\in\CC^\times\deq\CC\setminus\{0\}$ a non-zero complex number and $z_1,z_2\in\CC$ arbitrary complex numbers.
  
  First consider the special case $z_1=z_2=0$, $z_\times\in\CC^\times$ arbitrary.
  We denote by  $\mathbf R$ the corresponding subgroup of the little group $\Sg_{\latexchi_0}$.
  By writing
  \begin{equation*}
    z_\times = e^{-\ii {z}/2},\quad z\in\CC
    .
  \end{equation*}
  we see that $\mathbf R$ is isomorphic to $\Spin(2,\CC)$.
  Moreover, a straightforward computation (using the Weyl representation of the $\gamma$ matrices)
  shows that $\mathbf{R}$ is the subgroup of $\Spin(4,\CC)$
  of elements of the form
  \begin{equation*}
    R_0(z)\deq e^{\tfrac12\gamma_1\gamma_2 z},\quad z\in\CC
    ,
  \end{equation*}
  where $\gamma_1=\gamma(e_1)$, $\gamma_2=\gamma(e_2)$, and $e_0,e_1,e_2,e_3$
  is the standard basis of $\CC^4$.

  Now we turn to the special case where $z_\times=1$, $z_1,z_2\in\CC$ arbitrary.
  We denote by $\mathbf T$ the resulting subgroup of $\Sg_{\latexchi_0}$.
  Then it is straightforward to see that $\mathbf T$ is isomorphic to the group $\CC^2$
  considered as the group of translations in two complex dimensions.
  Moreover, using the Weyl representation of the gamma matrices, it is straightforward to show that
  $\mathbf T$ is the subgroup of $\Spin(4,\CC)$ of elements of the form
  \begin{equation*}
    T_0(z_1,z_2) \deq e^{ \frac12 (\gamma_3 + \gamma_0) . (x\gamma_1 + y\gamma_2)},
    \quad x \deq \tfrac12(z_2-z_1),\quad y\deq\tfrac{1}{2\ii}(z_1+z_2), \quad z_1,z_2\in\CC
    ,
  \end{equation*}
  which, in the Weyl representation, is equal to
  \begin{equation*}
    e^{ \tfrac12 (\gamma_3^{\text{Weyl}} + \gamma_0^{\text{Weyl}}) . (x\gamma_1^{\text{Weyl}} + y\gamma_2^{\text{Weyl}})}
    =
    \begin{psmallmatrix}
      1 & z_1 & 0 & 0\\
      0 & 1 & 0 & 0\\
      0 & 0 & 1 & 0\\
      0 & 0 & z_2 & 1
    \end{psmallmatrix}
    .
  \end{equation*}
  From this we see that a generic element of $s_0\in\Sg_{\latexchi_0}$ can be written as
  \begin{equation*}
    s_0(z_1,z_2,z) =  T_0(z_1,z_2)  R_0(z)
    ,
  \end{equation*}
  which in the Weyl representation becomes
  \begin{equation*}
    s_0^{\text{Weyl}}(z_1,z_2,z) =
    \begin{psmallmatrix}
      e^{-\frac{\ii}{2}z} & z_1 e^{\frac{\ii}{2}z} & 0                     & 0 \\
      0                 & e^{\frac{\ii}{2}z}     & 0                     & 0 \\
      0                 & 0                    & e^{-\frac{\ii}{2}z}     & 0 \\
      0                 & 0                    & z_2 e^{-\frac{\ii}{2}z} & e^{\frac{\ii}{2}z} 
    \end{psmallmatrix}
    .
  \end{equation*}
  with $z,z_1,z_2\in\CC$.
  A straightforward computation shows that we have the following multiplication law:
  \begin{equation}\label{eq:prod_rule}
    s_0(z_1,z_2,z) s_0(z_1',z_2',z') = s_0(z_1' + e^{-\ii z'} z_1,  z_2' +  e^{\ii z'} z_2,  z+z')
    ,
  \end{equation}
  with $z,z_1,z_2,z',z_1',z_2'\in\CC$.
  From this we obtain that $\mathbb S_{\latexchi_0}$ is isomorphic to the semidirect product
  \begin{equation*}
    \mathbb S_{\latexchi_0} \cong \CC^2 \rtimes \Spin(2,\CC)
    ,
  \end{equation*}
  where the semidirect product is the one associated with the standard action
  of $\Spin(2,\CC)$ on $\CC^2$ by conjugation.
  To see this, let $\CC^2$ be parameterized by elements $(x,y)\in\CC^2$,
  let $\CCliff(2)$ be the complex Clifford algebra generated by the Pauli matrices $\sigma_1,\sigma_2$,
  and consider $\Spin(2)\subset \CCliff(2)$ as the group of elements of the form
  $\varsigma = \exp\{\ii z \sigma_3/2\}$.
  If, as above, we consider the variables $z_1=-x+\ii y$, $z_2=x+\ii y$, then the action
  of $\Spin(2,\CC)$ on these variables is
  \begin{equation*}
    (z_1,z_2)\mapsto (e^{-\ii z}z_1 , e^{\ii z} z_2)
  \end{equation*}

  Then the natural action of $\Spin(2,\CC)$ on $\CC^2$ is given by
  \begin{equation*}
    x\ii\sigma_1+y\ii\sigma_2 \mapsto e^{\ii z \sigma_3/2} (x\sigma_1+y\sigma_2) e^{-\ii z\sigma_3/2}
    = (x \cos(z) + y \sin(z) )\sigma_1 + (-x\sin(z) + y\cos(z))\sigma_2
    .
  \end{equation*}
  From this and~\eqref{eq:prod_rule} 
  we see that $\Sg_{\latexchi_0}$ is indeed isomorphic to $\CC^2\rtimes\Spin(2,\CC)$.
  The proof now is complete. 
\end{proof}

\section{Irreducible unitary representations of $\ISpin(4,\CC)$ corresponding to non-zero ``complex mass''}\label{sec:non-zero-mass}
The general result in Theorem~\ref{th:general classification} gives,
together with Lemma~\ref{lem:orbit structure} and Lemma~\ref{lem:little group structure},
a full characterization of the irreducible unitary representations of $\ISpin(4,\CC)$
in terms of the irreducible unitary representations
of the corresponding little groups.
With reference to  Theorem~\ref{th:general classification},
we shall say that a representation 
of $\ISpin(4,\CC)=\CC^4\rtimes\Spin(4,\CC)$ is \textbf{associated}
to the character $\chi\in\hat\CC$
when the restriction of this representation to $\CC^4$
coincides with the character $\chi$.

In this section we restrict our attention to the irreducible unitary representations of
$\ISpin(4,\CC)$ associated to the character
$\latexchi_{\mathring v}\in\hat\CC$ relative to the vector
\begin{equation}\label{eq: standard vector}
  \mathring v\deq(z_m,0,0,0)\in\CC^4
  .
\end{equation}
The objective of this section is to describe,
in this special case,
the details of the realization of induced representations described in
Corollary~\ref{cor:induced_rep_2}.
This realization is particularly interesting because it generalizes the construction
done by Wigner in the case of the classical Poincar\'e group.
Moreover, in our opinion, thanks to this realization one can more directly relate
the Poincar\'e group with the four dimensional Euclidean group.
We believe this point to be relevant especially in the context Euclidean quantum field theory
and Wick rotation.

To realize the induced representations as defined in Corollary~\ref{cor:induced_rep_2},
we need three ingredients.
First,
we need to know the irreducible unitary representations
of the little group corresponding to the character $\latexchi_{\mathring v}$,
second,
we require a quasi-invariant measure on the orbit of $\latexchi_{\mathring v}$,
finally,
we need to chose a Wigner-Mackey embedding in the sense of our
Definition~\ref{def:Mackey-embedding}.
The first two ingredients are well known and we discuss them now, albeit only briefly.
In the Subsection~\ref{subsec:beta} below,
we shall describe three possible choices of Wigner-Mackey embeddings.
This will conclude our plan of describing explicitly and concretely the irreducible unitary representation of
$\ISpin(4,\CC)$ in the non-zero ``mass'' case.

For convenience we shall often identify, as in Section~\ref{sec:wigner}, $\hat\CC^4$ with $\CC^4$
and speak of the little group and orbit of $\mathring v\in\CC^4$ instead of the little group and orbit of
$\latexchi_{\mathring v}\in\hat\CC^4$.
\begin{parpar}[Little group of $\mathring v$.]
  It is straightforward to see that the little group relative to $\mathring v$ is given by
  \begin{equation*}
    \Sg_{\mathring v} = \{s\in\Spin(4,\CC)\st s\gamma_0s^{-1} = \gamma_0\},
  \end{equation*}
  where $\gamma_0\deq\gamma(e_0)$,
  with $e_0=(1,0,0,0)\in\CC^4$,
  and  $\gamma:\CC^4\rightarrow\CCliff(4)$
  is the Minkowski embedding in~\eqref{eq:Minkowski embedding}.
  By Lemma~\ref{lem:little group structure} we have that
  \begin{equation*}
    \Sg_{\mathring v} \cong {(\SL(2,\CC)\times\SL(2,\CC))}^{\text{diag}}\cong\SL(2,\CC)
    ,
  \end{equation*}
  where, as in the proof of Lemma~\ref{lem:little group structure},
  ${(\SL(2,\CC)\times\SL(2,\CC))}^{\text{diag}}\deq \{(A,A)\in\SL(2,\CC)\times\SL(2,\CC)\}$
  denotes the diagonal subgroup of $\SL(2,\CC)\times\SL(2,\CC)$.

  Hence the irreducible unitary representations of the little group of $\mathring v$
  are just the irreducible unitary representations of $\SL(2,\CC)$
  which are well-known.
  We refer e.g.\ to~\cite[Chapter 9, Theorem 2.5, p.\ 219]{taylorNoncommutativeHarmonicAnalysis1986},
  and reference therein,
  for the classification and description  
  of the irreducible unitary representations of $\SL(2,\CC)$.
\end{parpar}

\begin{parpar}[Invariant measure on the orbit of $\mathring v$.]
  By Lemma~\ref{lem:orbit structure} the orbit of $\mathring v$ is
  \begin{equation}\label{eq:orbit-iso}
    \mathscr O_{\mathring v} = \{s\gamma(\mathring v)s^{-1}\in\CC^4\st s\in\Spin(4,\CC) \} \cong z_m\SL(2,\CC)
    ,
  \end{equation}
  where $z_m$ is the ``complex mass'' as in~\eqref{eq: standard vector},
  and $z_m\SL(2,\CC)$ denotes the set of matrices of the form $z_m A$, for $A\in\SL(2,\CC)$.

  To realize the induced representation we need to fix a quasi-invariant measure $\mu_{\mathring v}$
  on the orbit $\mathscr O_{\mathring v}$. 
  As in~\eqref{eq:orbit-iso} we can identify $\mathscr O_{\mathring v}$ with $z_m\SL(2,\CC)$
  seen as the submanifold of $\CC^4$ of matrices $V$ with determinant $\det(V)=z_m^2$.
  From the isomorphism $\mathscr O_{\mathring v}\cong z_m\SL(2,\CC)$,
  it follows that we can choose a measure on $\mathscr O_{\mathring v}$ which is
  in fact \textit{invariant} under the action of $\ISpin(4,\CC)$ on $\mathscr O_{\mathring v}$,
  not just \textit{quasi}-invariant.
  This follows by the fact that $\SL(2,\CC)$ is unimodular and, therefore, admits a (both left and right invariant) Haar measure
  which is unique up to normalization.
  To make this invariant measure explicit, let us parameterize the matrices in
  $z_m\SL(2,\CC)$, up to a lower dimensional manifold (hence a zero measure set),
  as follows,
  \begin{equation*}
    V = \begin{psmallmatrix} (z_m^2+v_{12}v_{21})/v_{22} & v_{12}\\ v_{21} & v_{22} \end{psmallmatrix},
    \quad v_{12}, v_{21}, v_{22} \in\CC, \; v_{22}\ne0
    .
  \end{equation*}
  With this parameterization, we fix $\mu_{\mathring v}$ to be the following measure supported on $\CC^3$,
  \begin{equation*}
    \dd\mu_{\mathring v}(V) \deq \frac{1}{| v_{22}|^2}\dd v_{12}\dd v_{21}\dd v_{22}
    \dd \overline v_{12}\dd \overline v_{21}\dd \overline v_{22}
    ,
  \end{equation*}
  where $v=(v_{12},v_{21},v_{22})\in\CC^3$ and $\dd v_{12}\dd v_{21}\dd v_{22}
  \dd \overline v_{12}\dd \overline v_{21}\dd \overline v_{22}$
  denotes the Lebesgue measure on $\CC^3\cong\RR^6$.
  It is straightforward to show that the measure $\mu_{\mathring v}$ is invariant under
  the action of $\ISpin(4,\CC)$ on $\mathscr O_{\mathring v}$.
  Indeed, identifying $\mathscr O_{\mathring v}$ with $z_m\SL(2,\CC)$, the action of $\ISpin(4,\CC)$
  is given by $V\mapsto AVB^{-1}$, $V\in z_m\SL(2,\CC)$, $A,B\in\SL(2,\CC)$.
  It then suffices to note that the measure $\mu_{\mathring v}$, defined above,
  is both right and left invariant under the action of $\SL(2,\CC)$,
  as a straightforward computation shows.
  Regarding our choice of measure $\mu_{\mathring v}$ cf.\ e.g.\
  \cite[p.\ 68]{barut_theory_1986} and \cite[p.\ 13]{ruhl_lorentz_1970}.
  
  We remark that, since the measure $\mu_{\mathring v}$ is invariant,
  in particular, we have that the Radon-Nikodym derivative
  $\varrho_g$, $g\in\ISpin(4,\CC)$, in the formula~\eqref{eq:action orbit} of Corollary~\ref{cor:induced_rep_2},
  is constant and equal to $1$.
\end{parpar} 
This concludes the discussion regarding the first two ingredients
we need to construct the irreducible unitary representations of $\ISpin(4,\CC)$
relative the character $\latexchi_{\mathring v}$.
We will describe possible choices of a Wigner-Mackey embedding in the following subsection.
Assume one such a choice is made.
Then the irreducible unitary representations of $\ISpin(4,\CC)$
relative to $\latexchi_{\mathring v}$ take the following explicit form.
\begin{parpar}[Irreducible unitary representations corresponding to $\mathring v$.]\label{par:rep_z_m}
  Let $(\mathfrak H,\rho_{\mathring v})$ be a irreducible unitary representation of $\SL(2,\CC)$,
  where we have identified the little group of $\mathring v$ with $\SL(2,\CC)$.
  Similarly, identifying the orbit of $\mathring v$ with $z_m\SL(2,\CC)$,
  we fix on the orbit of $\mathring v$ the invariant measure $\mu_{\mathring v}=\mu_{z_m\SL(2,\CC)}$ described above.
  Then, the induced unitary representation $(\mathscr H,U)$, defined in Corollary~\ref{cor:induced_rep_2},
  is given, in this special case, by 
  \begin{equation*}
    \mathscr H = L^2(\mathscr O_{\mathring v},\mu_{\mathring v};\mathfrak H),
  \end{equation*}
  and, for $z\in\CC^4$, $v\in\mathscr O_{\mathring v}$,
  \begin{equation*}
    (U(z,s)f)(v) = \latexchi_{\mathring v}(z)\rho_{\mathring v}({\beta(v)}^{-1}s\beta(s^{-1}\cdot v) (f(s^{-1}\cdot v)),
  \end{equation*}
  where $\beta$ is a Wigner-Mackey embedding
  in the sense of our Definition~\ref{def:Mackey-embedding}.
\end{parpar}

\subsection{Wigner-Mackey embedding}\label{subsec:beta}
We would like to make the induced representation concrete in a way which parallels the classical Wigner case.
As remarked above, to do so we need to construct a Wigner-Mackey embedding $\beta$ in the sense
of Definition~\ref{def:Mackey-embedding}.
In the notation employed in the present section, this means that we are after a locally bounded measurable map
$\beta:\mathscr O_{\mathring v}\rightarrow \Spin(4,\CC)$, $v\mapsto\beta(v)$, $v\in\mathscr O_{\mathring v}$,
which satisfies
\begin{equation*}
  v = \beta(v)\cdot \mathring v, \quad v\in\mathscr O_{\mathring v},
\end{equation*}
where the $\cdot$ on the right hand side denotes, as before, the action of $\Spin(4,\CC)$ on $\CC^4$.
Employing the Minkowski embedding $\gamma:\CC^4\rightarrow\CCliff(4)$
defined in~\eqref{eq:Minkowski embedding}, this relation becomes
\begin{equation}\label{eq:beta}
  \gamma(v) = \beta(v) \gamma(\mathring v){\beta(v)}^{-1}, \quad v\in\mathscr O_{\mathring v}.
\end{equation}
As we will discuss in detail below,
in the classical Wigner case of $\ISpin^0(1,3)$, thanks to the isomorphism $\Spin^0(1,3)\cong \SL(2,\CC)$,
there is (up to a sign) a standard choice for this map which makes use of the polar decomposition of a matrix in $\SL(2,\CC)$.
For $\ISpin(4,\CC)$ the situation is somewhat more complicated.
To clarify this case we give three different concrete choices of a Wigner-Mackey embedding $\beta$.
\begin{parpar}[First realization.]
  Let us identify $\Spin(4,\CC)$ with $\SL(2,\CC)\times\SL(2,\CC)$. Then the little group
  $\Sg_{\mathring v}$ is isomorphic to ${(\SL(2,\CC)\times\SL(2,\CC))}^{\textrm{diag}}$.
  Moreover, the action of $\SL(2,\CC)\times\SL(2,\CC)$ on the orbit $\mathscr O_{\mathring v}\subset\CC^4$,
  defined in~\eqref{eq: action 2},
  takes the form 
  $\sigma((A,B).v)=A\sigma(v)B^{-1}$, with $(A,B)\in\SL(2,\CC)\times\SL(2,\CC)$ and $\sigma$ as in~\eqref{eq:map sigma}.
  
  Let us define the map $\pi_1:\SL(2,\CC)\times\SL(2,\CC)\rightarrow (\SL(2,\CC)\times\SL(2,\CC))^{\textrm{diag}}$
  which sends $(A,B)\mapsto(A,A)$, with $(A,B)\in\SL(2,\CC)\times\SL(2,\CC)$.
  We then have the decomposition
  \begin{equation*}
    (A,B) = (\id,BA^{-1})(A,A), \quad(A,B)\in \SL(2,\CC)\times\SL(2,\CC)
    .
  \end{equation*}
  We can now define
  \begin{equation}
    \label{eq:beta1}
    \beta_1(v) \deq (\id, (\sigma(v)/z_m)^{-1}),
    \quad v\in\mathscr O_{\mathring v}.
  \end{equation}
  It is straightforward to see that $\beta_1$ is a Wigner-Mackey embedding.
\end{parpar}
\begin{parpar}[Remark.]\label{rem:beta1}
  This construction has the advantage of taking a particularly simple form.
  Moreover the map $\beta_1$ is smooth.
  On the other hand, it has the disadvantage that,
  when we restrict from $\ISpin(4,\CC)$ to the classical Wigner case of $\ISpin^0(1,3)$,
  then $\beta_1$ does not restrict to a well defined Wigner-Mackey embedding there.
  Indeed, let us identify $\Spin^0(1,3)$ with the subgroup $\mathbf S_{1,3}$
  of $\SL(2,\CC)\times\SL(2,\CC)$
  of elements of the form $(A,A^{\ast-1})$, $A\in\SL(2,\CC)$.
  Moreover let $\mathscr O_m^\uparrow$ be the orbit of $(m,0,0,0)$ $m>0$, under $\mathbf S_{1,3}$.
  The restriction of the map $\beta_1$ is now
  \begin{equation*}
    \beta_1\rest_{\mathscr O_m^\uparrow}(p) = (\id,(\sigma(p)/m)^{-1})\not\in \mathbf S_{1,3},\quad p\in\mathscr O_m^\uparrow
    .
  \end{equation*}
  Hence the restriction of $\beta_1$ to this case does not define an embedding of $\mathscr O_m^\uparrow$
  into $\mathbf S_{1,3}$.
\end{parpar}
In the classical Wigner case, one defines the map
\begin{equation}
  \label{eq:WignerSquareroot}
  \beta_{\textrm{Wigner}}(p)\deq(\sqrt{\sigma(p)/m},(\sqrt{\sigma(p)/m})^{-1})\in\mathbf S_{1,3}, \quad p\in\mathscr O_m^\uparrow
  .
\end{equation}
Note that for $p\in\mathscr O_m^\uparrow$, $\sigma(p)$ is a positive-definite matrix, hence the square root is well defined,
unique up to a sign,
and $\beta_{\textrm{Wigner}}(p)$ in indeed in $\mathbf S_{1,3}$, $p\in\mathscr O_m^\uparrow$.
For reference we note that
\begin{equation*}
  \sqrt{\sigma(p)/m} = \frac{m \id + \sigma(p)}{\sqrt{2m(m+p_0)}},
  \quad
  p\in\mathscr O_m^\uparrow,
\end{equation*}
where $p_0$ denotes the first component of $p\in\mathscr O_m^\uparrow\subset\RR^{1,3}$.

In this Wigner case, the element $\beta_{\textrm{Wigner}}(p)\in\Sg_{1,3}\cong\Spin^0(1,3)$, $p\in\mathscr O_{\mathring v}$,
is usually referred to as a \textbf{standard boost} and the map $\beta_{\textrm{Wigner}}$ is denoted by the letter $L$
(cf.\ e.g.~\cite[p. 90]{thallerDiracEquation1992} or~\cite[p. 64, 68]{weinberg_quantum_1995}). 
In this paper we have chosen the letter $\beta$ for `boost'. 

The following two realizations give two possible extensions of this formula to the complexified case.
\begin{parpar}[Second realization.]
  As before, let us identify $\Spin(4,\CC)$ with $\SL(2,\CC)\times\SL(2,\CC)$.
  We employ the same notation as in the first realization above.
  We identify an element of the orbit $v\in\mathscr O_{\mathring v}$, with an element of $\SL(2,\CC)$ by $v\mapsto\sigma(v)/z_m\in\SL(2,\CC)$.
  Note that, as a manifold, $\SL(2,\CC)$ decomposes into a Cartesian product of $\mathscr P\times\SU(2)$ where $\mathscr P$ is the manifold of
  positive definite matrices with unit determinant.
  Explicitly, by polar decomposition we write $\sigma(v)/z_m=p(v)u(v)$ with $p(v)\deq\sqrt{\sigma(v)^\ast\sigma(v)}/|z_m|$ positive definite
  and $u(v)\in\SU(2)$.
  Then we define $\beta_2:\mathscr O_{\mathring v}\rightarrow \SL(2,\CC)\times\SL(2,\CC)$ by
  \begin{equation}
    \label{eq:beta2}
    \beta_2(v) \deq (\sqrt{p(v)},u(v)^{-1}(\sqrt{p(v)})^{-1}),
    \quad v \in \mathscr O_{\mathring v}
    .
  \end{equation}
  A straightforward computation shows that $\beta_2$ satisfies  \eqref{eq:beta}.
\end{parpar}
\begin{parpar}[Remark.]
  Note that the map $\beta_2$ is not smooth (because $A\mapsto |A|\deq\sqrt{A^\ast A}$, $A\in\SL(2,\CC)$, is not smooth)
  but it is still continuous.
  Moreover, restricting from $\Spin(4,\CC)$ to $\Spin^0(1,3)$ as in the remark~\ref{rem:beta1} above,
  we see that the restriction of $\beta_2$ to the orbit $\mathscr O_m^\uparrow$
  coincides with the standard Wigner boost $\beta_{\text{Wigner}}$.
\end{parpar}
\begin{parpar}[Third realization.]
  Consider the isomorphism $\mathscr O_{\mathring v}\cong\SL(2,\CC)$ given by
  $v\mapsto\sigma(v)/z_m$, with $v\in\mathscr O_{\mathring v}$.
  Then let us define
  \begin{equation}
    \label{eq:beta2}
    \beta_3(v) \deq
    \begin{cases}
      \beta^+(v), & v\in \mathscr O_{\mathring v}\setminus{-\mathring v},\\
      \Omega, & v = -\mathring v,
    \end{cases}
  \end{equation}
  where
  \begin{align*}
    \beta^+(v) &\deq \frac{z_m\id + \gamma(v)\gamma_0}{\sqrt{2z_m(z_m+v_0)} },
                 \quad v\in\mathscr O_{\mathring v}, \quad \Re(v_0/z_m)\ge 0,
  \end{align*}
  and where $\Omega$ is the ``volume form'' defined in~\eqref{eq:volume_form}.
\end{parpar}
\begin{parpar}[Remark.]\label{rem:discontinuity}
  The map $\beta_3$, like $\beta_2$, 
  restricts to the map $\beta_{\text{Wigner}}$ (cf.\ remark~\ref{rem:beta1}) when we restrict from $\Spin(4,\CC)$ to $\Spin^0(1,3)$.
  A disadvantage of $\beta_3$ is that it is clearly discontinuous.
  Let us clarify the particular form we chose for the map $\beta_3$.
  First note that, in the Weyl representation of the Clifford algebra $\CCliff(4)$,
  we have that
  $\beta^+(v)$
  is a block diagonal matrix, with $v\in\mathscr O_{\mathring v}$. Hence the map $\beta^+$
  defines (by restricting to, say, the first block of the matrix)
  a map $j^+:\mathscr O_{\mathring v}\setminus\{-\mathring v\} \rightarrow\SL(2,\CC)$,
  given by
  \begin{equation}
    \label{eq:discontinuous_squareroot_SL}
    j^+: v \mapsto j^+(v) \deq \frac{z_m\id + \sigma(v)}{\sqrt{2z_m(z_m+ v_0)}},
    \quad
    v\in\mathscr O_{\mathring v}\setminus\{-\mathring v\}
    . 
  \end{equation}
  Note that this formula extends formula~\eqref{eq:WignerSquareroot} in remark~\ref{rem:beta1}
  to the case $v\in\mathscr O_{z_m}\setminus{-\mathring v}$.
  A straightforward computation shows that
  \begin{equation*}
    \sigma(v)/z_m = j^+(v)^2,
    \quad
    v\in\mathscr O_{z_m}\setminus\{-\mathring v\}
    .
  \end{equation*}
  Hence the  $j^+(v)\in\SL(2,\CC)$ is a ``square root'' of $\sigma(v)/z_m\in\SL(2,\CC)$
  for $v\in\mathscr O_{z_m}\setminus\{-\mathring v\}$.
  Note that for $v=-\mathring v$
  both the numerator and the denominator
  in right hand side of~\eqref{eq:discontinuous_squareroot_SL} go to zero.
  The limit of $v\rightarrow\mathring v$ in~\eqref{eq:discontinuous_squareroot_SL}
  is not well defined. In fact, it
  depends on the chosen ``direction'', i.e.\ on the choice of one parameter subgroup connecting $\mathring v$ with $-\mathring v$.
  Indeed, let 
  $\hat{\mathbf u}\deq(u_1,u_2,u_3)$ be an element of the two-sphere
  $S^2\subset\RR^3$, that is $\hat u_1, \hat u_2, \hat u_3\in\RR$, $\hat u_1^2+\hat u_2^2+\hat u_3^2=1$.
  For any such ``direction'' $\hat{\mathbf u}$ we define 
  $u(\theta)\in\mathscr O_{\mathring v}$, $\theta\in[0,\pi)$ by
  \begin{equation*}
    \sigma(u(\theta)) = \Big(\cos(\theta)\id+\sin(\theta)\sum_{j=1}^3\ii\hat u_j\sigma_j\Big)\sigma(\mathring v)
    .
  \end{equation*}
  Then we have
  \begin{equation*}
    j^+(u(\theta))=
    \frac{z_m\id + z_m\big(\cos(\theta)\id +\sin(\theta)\sum_{j=1}^3\ii\hat u_j\sigma_j\big)}
    {\sqrt{2z_m(z_m-z_m\cos(\theta))}},
    \quad \theta\in[0,\pi)
    .
  \end{equation*}
  A straightforward computation shows that, in the limit $\theta\rightarrow\pi$, we have
  \begin{equation*}
    \lim_{\theta\rightarrow\pi}j^+(u(\theta))= z_m\ii\sum_{j=1}^3 \hat u_j\sigma_j
    .
  \end{equation*}
  Hence the limit depends on the direction $\hat{u}=(u_1,u_2,u_3)\in S^2\subset\RR^3$.
  
  We can imagine pictorially the discontinuity at $-\id$
  of the square root $j^+(v)$ of $\sigma(v)/z_m$, $v\in\mathscr O_{\mathring v}$ as follows.
  Since $-\id$ is in the image of the exponential map $\exp:\mathfrak{sl}(2,\CC)\rightarrow\SL(2,\CC)$
  let us restrict our attention to elements
  $A\in\SL(2,\CC)\setminus\{-\id\}$ which are in the image of the exponential map.
  Any such element $A$
  determines a unique one-parameter subgroup of $\SL(2,\CC)$.
  We imagine to join $A$ with the identity with a path following this one-parameter subgroup.
  Then the square root of $A$ is given by the element on this path which sits half way
  between $\id_2$ and $A$.
  With this picture in mind, it becomes clear that the square root of
  $-\id=-\sigma(\mathring v)/z_m=-\id_2\in\SU(2)\subset\SL(2,\CC)$
  is highly degenerate, indeed any $\mathbf x\in S^2\subset\RR^3$, $\|\mathbf x\|_{\RR^3}=1$,
  defines a matrix $\ii\boldsymbol \sigma\cdot \mathbf x\in\SU(2)$ the square of which is $-\id_2$.

  As a consequence we see that the extension of $j^+(v)$ to $v\in\SL(2,\CC)$ is not unique and
  any possible extension will be discontinuous.
  We remark this also in relationship to~\cite{osterwalder_euclidean_1973-2}
  putting in evidence that the formulae specified there (cf.\ formula (3,6) of~\cite{osterwalder_euclidean_1973-2})
  do not hold for
  $p=(-m,0,0,0)$.
\end{parpar}
\begin{parpar}[Remark.]
  In our definition of $\beta_3$ we have chosen to define the map $\beta_3$ at the point
  $-\mathring v$ to be $\beta_3(-\mathring v)=\Omega$.
  Another option would be to ``cut'' $\SL(2,\CC)$ along the $S^2$ described in the remark above
  and to define a different square root on each ``hemisphere''. 
  Explicitly we could define $\beta'_3:\mathscr O_{\mathring v}\rightarrow\Spin(4,\CC)$ to be
  \begin{equation}
    \label{eq:beta2-prime}
    \beta'_3(v) \deq
    \begin{cases}
      \beta^+(v), & \Re(v_0/z_m)\ge 0, \quad v\in\mathscr O_{\mathring v},\\
      \beta^-(v), & \Re(v_0/z_m)< 0, \quad v\in\mathscr O_{\mathring v},
    \end{cases}
  \end{equation}
  where $v_0$ is the first component of $v\in\mathscr O_{\mathring v}\subset\CC^4$,
  $\Re(v_0/z_m)$ the real part of $v_0/z_m$, and 
  \begin{align*}
    \beta^+(v) &\deq \frac{z_m\id + \gamma(v)\gamma_0}{\sqrt{2z_m(z_m+v_0)} },
                 \quad v\in\mathscr O_{\mathring v}, \quad \Re(v_0/z_m)\ge 0,
    \\
    \beta^-(v) &\deq \Omega\beta^+(-v), \quad v\in\mathscr O_{\mathring v},
                 \quad \Re(v_0/z_m)< 0,
  \end{align*}
  where, as above, $\Omega=\ii\gamma_5$.
\end{parpar}

\section{Conclusion}
In the last section, we have given an explicit realization of the irreducible unitary representations of $\ISpin(4,\CC)$
relative to the character $\latexchi_{\mathring v}$, $\mathring v=(z_m,0,0,0)$, $z_m\in\CC\setminus0$.
We note that our analysis of $\ISpin(4,\CC)$ applies effectively unchanged
to the Euclidean case, where one considers the group
$\ISpin(4)\deq\RR^4\rtimes\Spin(4)$ acting on $\RR^4$ (here $\Spin(4)$ denotes the Spin
group on $\RR^4$ with respect to the standard Euclidean metric).
Moreover, having treated the complexified case, we can relate more directly the standard
Wigner construction of the representations of $\ISpin^0(1,3)$ with
the representations of the group $\ISpin(4)$.
The representations of the group $\ISpin(4)$, and their relation with those of $\ISpin^0(1,3)$,
are relevant, for example, when one passes
from the Minkowski quantum field theory to the Euclidean one.
It is common in this context to employ what we have called a ``Wigner-Mackey embedding'' 
to write down explicit formulas and computations.
For example, as noted at the end of Remark~\ref{rem:discontinuity}, an embedding of this type
was employed in~\cite[formula (3,6)]{osterwalder_euclidean_1973-2}.
Here we have tried to give a general description of these embeddings,
while also giving concrete, explicit examples.

\section{Acknowledgments}
We thank Prof.\ Dr.\ Sergio Albeverio for numerous comments and very inspiring discussions during the development of this work. 
Moreover we would like to express our deep
gratitude to him for his support and encouragement during these difficult pandemic times.


\printbibliography

\end{document}%